\documentclass[aps,prx,reprint,twocolumn,longbibliography,superscriptaddress, floatfix]{revtex4-2}
\usepackage{graphicx} 
\usepackage{bm}
\usepackage{amssymb} 
\graphicspath{ {images/} }
\usepackage[english]{babel}
\usepackage{graphicx}
\usepackage{amsfonts}
\usepackage{amssymb}
\usepackage{amsmath}
\usepackage{amsthm}
\usepackage[citecolor=blue, colorlinks=true, urlcolor=blue, linkcolor=blue]{hyperref}
\usepackage{txfonts}

\begin{document}

% affiliation
\newcommand{\TUM}{\affiliation{Technical University of Munich, TUM School of Natural Sciences, Physics Department, 85748 Garching, Germany}}
\newcommand{\MCQST}{\affiliation{Munich Center for Quantum Science and Technology (MCQST), Schellingstr. 4, 80799 M{\"u}nchen, Germany}}
\newcommand{\WUSTL}{\affiliation{Department of Physics, Washington University in St. Louis, 1 Brookings Dr., St. Louis MO 63130, USA}}

\author{Caterina Zerba}
\TUM \MCQST
\author{Alexander Seidel}
\WUSTL
\author{Frank Pollmann}
\TUM \MCQST
\author{Michael Knap}
\TUM \MCQST

\def\papertitle{{Squeezing}}

\def\papertitle{{Validation of Fractional Quantum Hall experiments via anomalous relaxation}}

\def\papertitle{{Emergent Fracton Dynamics as a Probe of Fractional Quantum Hall Physics}}

\def\papertitle{{Emergent Fracton Hydrodynamics in the Fractional Quantum Hall Regime of Ultracold Atoms}}

\def\papertitle{{Emergent Fracton Hydrodynamics of Ultracold Atoms in Partially Filled Landau Levels}}

\title{Emergent Fracton Hydrodynamics of Ultracold Atoms in Partially Filled Landau Levels}

\date{\today}

\begin{abstract}

The realization of synthetic gauge fields for charge neutral ultracold atoms and the simulation of quantum Hall physics has witnessed remarkable experimental progress. 
Here, we establish key  signatures of fractional quantum Hall systems in their non-equilibrium quantum dynamics. We show that in the lowest Landau level the system generically relaxes subdiffusively.
The slow relaxation is understood from  emergent conservation laws of the total charge and the associated dipole moment that arise from the effective Hamiltonian projected onto the lowest Landau level, leading to subdiffusive fracton hydrodynamics. We discuss the prospect of rotating quantum gases as well as ultracold atoms in optical lattices for observing this unconventional relaxation dynamics.

\end{abstract}

\maketitle

\section{Introduction}

In solid state systems, fractional quantum Hall states typically arise under extreme conditions of ultra-low temperatures, large magnetic fields, and strong interactions~\cite{Tsui1982, Laughlin83}. These exotic states of matter with topological order exhibit charge fractionalization and anyonic excitations. 
Recent experimental and technical advances enabled the study of quantum Hall physics with synthetic quantum systems as well, which offer new routes to probe these exotic states. Synthetic systems of ultracold atoms are charge neutral and hence do not minimally couple to a magnetic field, instead artificial gauge fields have to be engineered. One way to realize artificial gauge fields for ultracold quantum gases in the continuum is by using the equivalence of neutral particles under rotation and charged particles in a magnetic field~\cite{Matthews1999,Madison2000, Abo-Shaeer2001,Schweikhard2004,Zwierlein2005}. By properly designing the rotating trapping potential, recent work has demonstrated the geometric squeezing of a condensate into the Lowest Landau Level (LLL)~\cite{Fletcher2021Jun,Mukherjee2022Jan, Yao2023, Crepel2023}. Moreover, few-body Laughlin states have been created by rotating fermions in an optical tweezer~\cite{Lunt2024}.
Artificial magnetic fields have also been realized for ultracold quantum gases by spatially dependent optical couplings~\cite{Lin2009Dec,Lin2009,Chalopin2020Oct} and in optical lattices by laser dressing~\cite{Struck2012, Aidelsburger2013, Miyake2013, Jotzu2014, Atala_2014, Aidelsburger2015, Stuhl2015, Mancini2015, Flaschner2016}. The latter approach enabled the observation of interaction-induced propagation of chiral excitations~\cite{Tai2017} as well as the preparation of fractional quantum Hall (FQH) states in the few-body limit~\cite{Leonard2023}. Furthermore, FQH states have been investigated also with photons in circuit QED and cavities~\cite{Roushan2017, Clark_2020, Wang2024May}.
%Nonetheless, adiabatically preparing fractional FQH ground states in the many-body limit and probing their intricate properties remain central open challenges. 

Key experimental challenges in the field are to realize the Hamiltonian with synthetic gauge fields, to cool the system to the LLL with a finite density of particles, and to prepare the FQH ground state. In this work, we propose to study the non-equilibrium relaxation dynamics to identify key signatures of the partially filled LLL at finite energy densities. Recently, it has been shown that quenching the effective mass tensor excites the dipole and graviton modes of FQH states~\cite{Liu2018,Fremling2018Oct, Liu2021}. These quench protocols are initialized with the ground state of the system. By contrast, here, we will avoid the challenge of preparing the interacting ground state and consider relaxation at finite energy densities.

It is the common anticipation that generic quantum many-body systems at finite energy densities relax diffusively in time. How quantum systems quenched to the LL relax is an interesting question, as they are subjected to strong constraints. In this work, we show that the relaxation dynamics is strongly modified in these systems and becomes subdiffusive. In particular, we explore the unique signatures of fractionally occupied states in the LLL by investigating the relaxation of density excitations in the far-from equilibrium dynamics of a finite energy-density state in the LLL. 
It is known that interacting particles in the LLL are described by effective %one-dimensional 
models which conserve both the global charge as well as the associated center of mass (or equivalently the dipole moment)~\cite{Haldane1983,Haldane1985,Trugman1985,Bergholtz2005, Alex2005, Nakamura2012, Bergholtz2006, Rezayi1994, Bergholtz2008,Moudgalya2020,Bergholtz2005, Nachtergaele2020Nov}. Due to these effective conservation laws, the mobility of the particles is constrained. Following the predictions of fracton hydrodynamics, that emerges in systems with both charge and dipole conservation~\cite{Gromov2020, Feldermeier2020, Guardado2020}, we show that these systems exhibit slow relaxation dynamics due to interaction-induced intra-Landau-level scattering processes.
This unconventional relaxation dynamics thus  characterizes the partially-filled LLL, which can be directly probed in experimental platforms of ultracold atoms. 

Our results are organized as follows. In Sec.~\ref{LLL} we introduce the effective model for the LLL; in Sec.~\ref{dynamicsLLL} we address effective high-temperature properties of states in the LLL by analyzing the emergent fracton hydrodynamics of interacting particles; we find anomalous relaxation dynamics of the density-density autocorrelation function as well as the bipartite fluctuations of the particle number in a subregion of the system. In Sec.~\ref{Ultracold} we discuss  experimental platforms for realizing FQH physics with ultracold atoms, focusing both on rotating  quantum gases and on laser-dressed atoms in optical lattices. In both cases, we  analyze the full, unprojected dynamics and determine the conditions for conserving the dipole moment in real space. An outlook and discussions are provided in Sec.~\ref{outlook}. 

\section{Effective model for the Lowest Landau Level}\label{LLL}

\begin{figure}
\includegraphics[width=0.98\columnwidth]{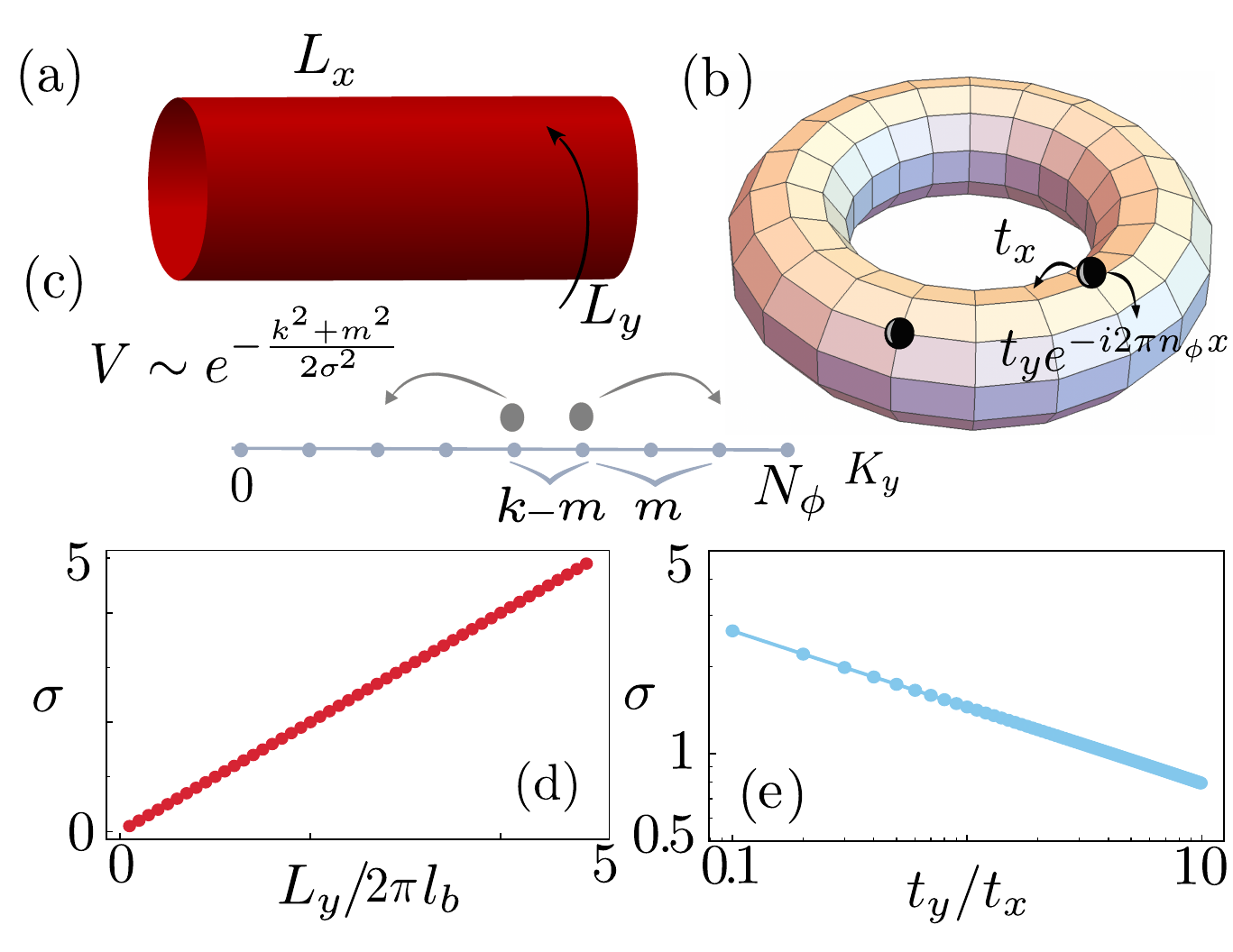}
\caption{\textbf{Projection onto the lowest Landau level.} (a) Illustration of the continuum model on a cylinder. (b)  Illustration of the Harper-Hofstadter model on a torus in the Landau gauge; flat bands are realized for specific lattice geometries, as discussed in the main text.
(c) When projecting onto the lowest band, the contact interaction gives rise to center-of-mass (i.e., dipole-moment) conserving hopping terms, whose matrix elements decay with typical length scale $\sigma$. (d) For the continuum model the typical length scale $\sigma$ depends only on the ratio $L_y/l_b$. (e) For the interacting Harper-Hofstadter model $\sigma$ can be tuned by changing the hopping ratio $t_y/t_x$ ($\sigma$ is numerically evaluated for system size of $N_x=44$ and $N_y=22$).}
\label{fig:1}
\end{figure}

We will now consider particles confined to a two-dimensional plane subjected to a strong perpendicular magnetic field, such that Landau levels form. When considering a finite energy-density state within the LLL, their dynamics remains restricted to the LLL when the interaction energy between particles is small compared to the energy gap between Landau levels. Consequently, relaxation is governed by interactions projected onto the LLL. As the matrix elements of the projected interactions introduce a notion of distance between different quantum numbers, the relaxation effectively occurs in a one dimensional lattice. We will now derive the projected models both in the continuum and on a lattice. In both cases, we demonstrate that the system possesses an emergent dipole-conserving dynamics within this effective one-dimensional lattice.

We first focus on the continuum, where the single-particle Hamiltonian that describes the system is $H_0=\frac{1}{2m}(\bold p-\bold A)^2 $. Here, $(\bold p-\bold A)$ is the generalized momentum in the presence of the magnetic field, and $\bold A =B \, (0,x)$ is the vector potential in the Landau gauge. 
The properties of the system are determined by the magnetic length $l_b=\sqrt{\frac{\hbar}{e B}}$, the number of magnetic fluxes $N_\phi = \frac{L_x L_y}{2 \pi l^2_b}$, and the cyclotron frequency $\omega_c=\frac{eB}{m}$. Assuming an infinite cylinder geometry with periodic boundary conditions in the $y$-direction, Fig.~\ref{fig:1}(a), the basis that diagonalizes the single-particle Hamiltonian is~\cite{Haldane1983,Haldane1985,Trugman1985,Moudgalya2020}, 
$\psi_{\gamma}^{m}(\bold r)=\frac{1}{\sqrt{{\pi}\gamma!2^\gamma L_y}}H_\gamma\bigg(\frac{x-x_m}{l_B}\bigg)e^{-\frac{1}{2}(\frac{x-x_m}{ l_b})^2}e^{ik_m y}$,
where $H_n$ are the Hermite polynomials, $x_m= 2 \pi m\frac{l^2_b}{L_y}$ with $m \in [0, N_\phi-1]$ determine the center of the wave function, and $k_m=\frac{2 \pi m}{L_y}$ is the momentum in $y$-direction. The associated single-particle eigenvalues are $E_\gamma^m=\hbar \omega_c (\gamma+\frac{1}{2})$. Assuming that particles initially occupy only one Landau level, density-density interactions $V=\sum_{i,j}V_{i,j}n_i n_j$ of magnitude much smaller than the inter-level spacing $\omega_c$, give rise to an effective dynamics that is confined to the same Landau level. The  Hamiltonian is obtained by projecting the interacting Hamiltonian $H=H_0+V$ onto the Landau level~\cite{Moudgalya2020} 
\begin{multline}\label{eq:LLLcont}
  P_0HP_0=\sum_{q=0}^{N_\phi-1} \sum_{p=0}^{\frac{N_\phi}{2}-1}\sum_{k=p+1}^{\frac{N_\phi}{2}} (V^{\text{con}}_{k,p} b^\dagger_{q}b^\dagger_{q+k+p}b_{q+k}b_{q+p} + \text{h.c.}),
\end{multline}
where $p,q,k \,\in [0, N_\phi-1]$  are integers that label the $y-$component of the momenta in the  LLL, and $N_\Phi$ coincide with the number of magnetic cells in the $y-$direction. The effective model is one-dimensional and consists of squeezing terms in momentum space, that conserve both the total charge $\tilde{\mathcal{N}} =\sum_q b^\dag_q b_q$ as well as the dipole moment in the LLL $\tilde{\mathcal{P}} = \sum_q q\, b^\dag_q b_q$; Fig.~\ref{fig:1}(c). The squeezing motion can be understood from the conservation of total momenta in the $y$-direction, since the momentum labels the sites of the  one-dimensional lattice. 
When assuming contact interactions in the unprojected system and considering the LLL, the matrix elements of the projected interaction  depends only on the overlap in the $x$-component of the eigenstates in the LLL. As such, they can be tuned by changing the parameters in the Hamiltonian. For our model, $V^{\text{con}}_{k,p}\propto  \exp[{-\frac{k^2+p^2}{2\sigma^2}}]/[(1+\delta_{p,0})(1+\delta_{k,{N_y}/{2}})]$ \cite{Moudgalya2020,Ortiz13}, where the typical length scale $\sigma = L_y/{2}\pi l_b$ and the denominator avoids double counting. The effective matrix elements $V^{\text{con}}_{k,m}$ of the projected Hamiltonian can be tuned by  changing the magnetic length $l_b$, or equivalently the thickness of the cylinder, see Fig.~\ref{fig:1} (d). As $l_b$ increases the elements of the potential decay faster, and in turn a slower dynamics in the LLL arises. This is directly obtained from the dependence of the single-particle eigenstates on the magnetic length. 

An equivalent effective model can be obtained for interacting bosons in a 2D lattice with magnetic field, described by the interacting
 Harper-Hofstadter model \begin{multline}\label{eq:1}
H=\sum_{x,y} [-t_x a^{\dagger}_{x+1,y}a_{x,y} - t_y e^{-i 2 \pi n_{\phi}x}a^{\dagger}_{x,y+1}a_{x,y} + \text{h.c.}] \\+ \frac{U}{2}\sum_{x,y}n_{x,y}(n_{x,y}-1)\equiv H_0+V.
\end{multline}
where $n_\phi=P/Q$ is the number of fluxes per unit cell, $P,Q$ integers and coprime. The non-interacting system $H_0$ is invariant under magnetic translations, 
\begin{align}\label{eq:2-3}
&T_x=\sum_{x,y} e^{i 2 \pi n_{\phi}y}a^{\dagger}_{x+1,y}a_{x,y} \nonumber\\
&T_y=\sum_{x,y} a^{\dagger}_{x,y+1}a_{x,y} ,
\end{align}
as $[H_{0}, T_y]=[H_{0}, T_x]=0$; since moreover $[T_x^Q, T_y]=0$, the three operators $H_{0}, T_x^Q $and $T_y$ can be diagonalized simultaneously. In contrast to the continuum,  the bands of the Harper-Hofstadter model for $U=0$ are not flat for generic parameters. This leads to additional terms in the effective Hamiltonian projected onto the lowest band compared to Eq.~\eqref{eq:LLLcont}. Interestingly, it has been found that it is possible to define geometric constraints on the lattice such that for finite-size systems the resulting bands are flat, effectively one-dimensional, and topologically non-trivial~\cite{Scaffidi2014}. Due to the finite system this result does not contradict certain no-go theorems on the absence of flat topological bands in lattice systems with locality~\cite{Chen2014}.  

Non-trivial, one-dimensional flat bands are obtained when the lattice has  $N_x=Q$ sites in the $x$-direction with $Q=\alpha N_y$,  $\alpha$ integer, and $N_y$ sites in the $y$-direction, that in this direction corresponds also to the number of magnetic unit cells.  Then the form of the eigenfunctions is $\psi_{K_y,n}(x,y)=u_{K_y, n }(x)e^{-i\frac{2\pi}{N_y}K_yy} /\sqrt(N_y)$ where $K_y \in [0,...N_y-1]$ are quantum numbers of the operator $T_y$, $n$ defines the energy level and $u_{K_y, n}(x)$ are periodic functions in the $x$-direction $u_{K_y,n}(x+Q)=u_{K_y, n}(x)$. For these parameters the Hofstadter equations 
\begin{multline}\label{hofstd}
 - E_n u_{K_y,n}(x)=2 t_y \cos\left[2\pi\left( \frac{K_y}{N_y}+\frac{P}{Q}x\right)\right]u_{K_y,n}(x)+\\+ t_x \big[ u_{K_y,n}(x+1)+u_{K_y,n}(x-1) \big]
\end{multline}
are the same as the momenta are changed, up to a translation in the $x-$direction, guaranteeing flat bands~\cite{Scaffidi2014}.  
Similarly to the continuum case, one can interpret each flat band as a one-dimensional lattice whose sites are identified by the quantum numbers $K_y$. Moreover, due to the lattice geometry the functions $u_{K_y, n }(x)$ for different $K_y \not= K_y'$ are the same apart from a translation in the $x$-direction.
Projecting the on-site interaction with a projector $P_0$ onto the LLL, again leads to an effective one-dimensional dipole-conserving model
\begin{equation}\label{eq:projH}
  P_0HP_0=\sum_{q=0}^{N_y-1} \sum_{p=0}^{\frac{N_y}{2}-1}\sum_{k=p+1}^{\frac{N_y}{2} }(V_{k,p}^\text{lat} b^\dagger_{q}b^\dagger_{q+k+p}b_{q+k}b_{q+p} + \text{h.c.}).
\end{equation}
We present the details of this calculation in App.~\ref{ProjectionLLL}. The matrix elements $V_{k,p}^\text{lat}\sim \exp[{-\frac{k^2+p^2}{2\sigma^2}}]/[(1+\delta_{p,0})(1+\delta_{k,{N_y}/{2}})]$ are approximately scaling as a Gaussian, similarly to the continuum, and are controlled by the width of the single-particle orbitals that are tunable by the ratio of the hopping matrix elements $t_y/t_x$; see Fig.~\ref{fig:1}{(e)}. 
The dipole squeezing motion is again suppressed with increasing overall separation of the hopping processes via a Gaussian shaped potential $V_{k,m}$.
Even thought the precise geometric constraints may be challenging to realize experimentally, this flat-band model severs as a controlled starting point for our analysis. Consequences of relaxing the flat-band conditions are discussed in App.~\ref{NonFlatbands}.
From this model, the continuum limit is recovered by sending $Q\to \infty$; in this limit, up to exponentially small corrections, the number of sites in the effective one-dimensional model Eq.~\eqref{eq:projH} corresponds to the number of magnetic fluxes~\cite{Scaffidi2014, Sorensen2005,Hafezi2007,Harper2014}.

Thus, for both models analyzed in this section, we have obtained effective dipole conserving dynamics in a one dimensional lattice. For both cases, the mechanism that allows for the emerging dipole conservation in the LLL is the conservation of one of the components of momentum. We will now show how this unconventional conservation law determines subdiffusive behavior in the LLL and how the geometry of the model can be used to tune dynamical properties. 

\section{Dynamics in the Lowest Landau Level}\label{dynamicsLLL}
We now investigate the far-from equilibrium relaxation dynamics in the LLL. The effective models projected onto the LLL in Eqs.~\eqref{eq:LLLcont} and \eqref{eq:projH} consists of a sum of squeezing terms with tunable coefficients which conserves both the total charge and the total dipole moment on a one-dimensional lattice. 

For concreteness, we focus here on the projected Harper-Hofstadter model. The results apply similarly for the continuum model, as only the matrix elements of the squeezing terms in Eq.~\eqref{eq:LLLcont} and Eq.~\eqref{eq:projH} slightly differ. The matrix elements are determined by the typical length scale $\sigma$ of the Gaussian envelop function. For $\sigma \lesssim 1$ the so-called thin torus limit is realized and the dominant contribution to the Hamiltonian are on the one hand the density-density interactions ${n}_{q} {n}_{q+1}$, where ${n}_{q}=b^\dagger_q b_q$ and on the other hand the shortest-distance squeezing term $b_{q-1}^\dag b_q b_{q+1} b^\dag_{q+2}$. When considering only this contribution and a maximum local occupation of one boson per site, the system exhibits strong Hilbert space fragmentation: even when fixing the charge and dipole quantum number, the Hilbert space fragments into exponentially many Krylov subspaces, each of which is parametrically small compared to the total dimension of the Hilbert space sector~\cite{Sala2020,Rakovszky2020,Khemani2020,Morningstar2020}. As a consequence, correlation functions do not decay in time. Longer-range terms in the Hamiltonian give rise to weak Hilbert space fragmentation and restore ergodicity; for a given charge and dipole density an excitation relaxes~\cite{Sala2020}. Even though contributions to the Hamiltonian are exponentially suppressed at larger distances, they still exist and thus lead, in the thermodynamic limit, to ergodic relaxation dynamics at a certain time scale. 

 \begin{figure}
  \centering
  \includegraphics[width=\columnwidth]{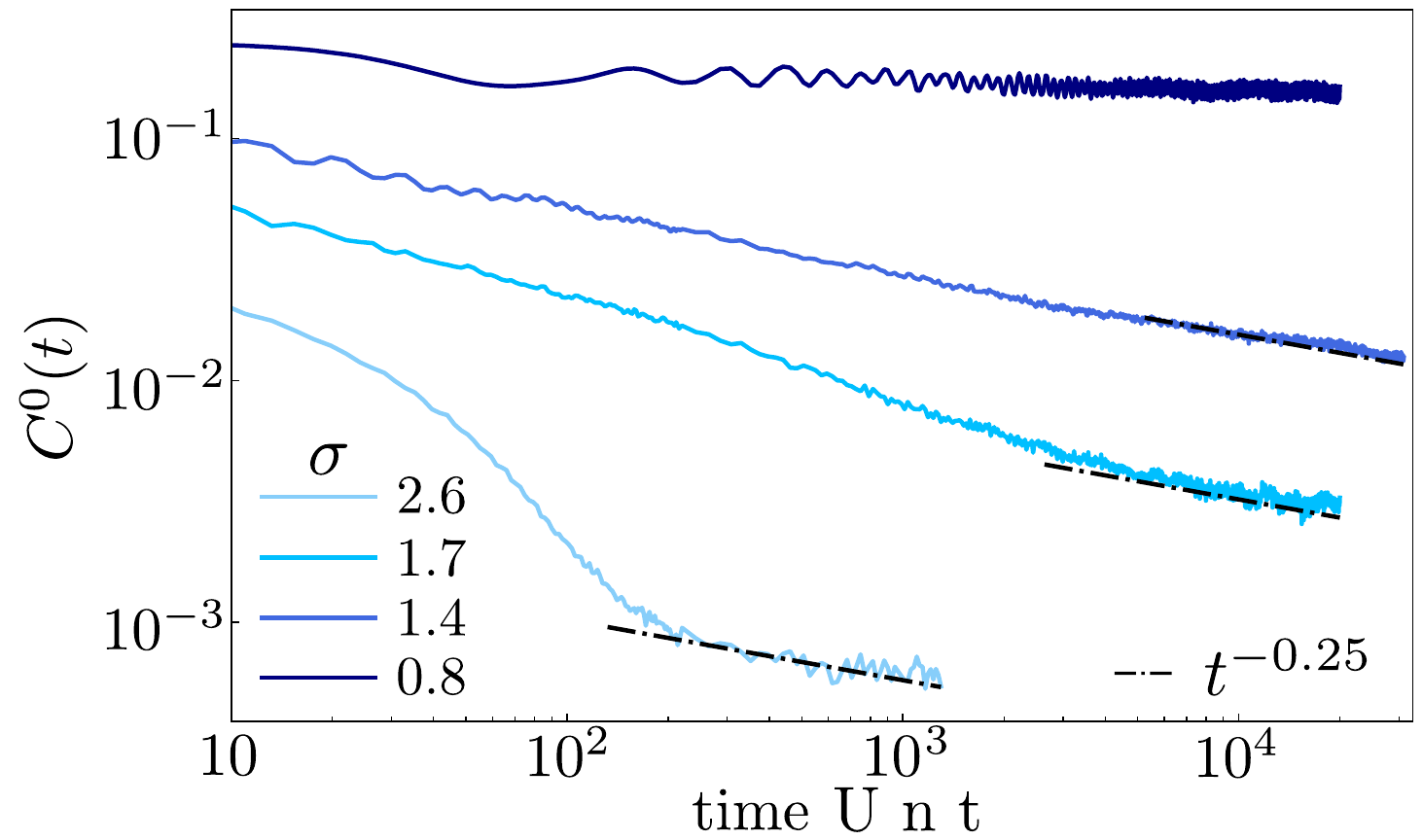}
  \caption{\textbf{Density-density autocorrelation function.} Time evolution of the autocorrelation function $C^0(t)=\langle\Delta n_0(t) \Delta n_0(0)\rangle$ for different typical interaction length scales $\sigma$, where $\Delta n_0= b^\dagger_0b_0- \langle b^\dagger_0b_0\rangle{}_{\rho_0}$ and $\langle...\rangle{}_{\rho_0}$ is the average over the chosen sector of the density operator. 
  }
  \label{fig:3}
\end{figure}

Our goal is now to study this crossover at finite times by tuning the typical interaction length scale $\sigma$. 
We numerically compute the time evolution of the system in an infinite temperature state of the LLL using exact diagonalization. The initial state serves as a generic finite energy-density state in the LLL. % and is obtained by randomly sampling product states in a fixed particle number and dipole sector of the LLL, with the same probability, %The initial state is thus given by the ensemble where all the states in the sector appear with the same probability. 
%as the evolution of the full ensemble is computationally intractable for large systems. 
While the short-time dynamics depends on the details of the initial state, the emergence of hydrodynamics relies only on the symmetries of the problem. Similar late-time dynamics to the one presented here, will therefore also arise when a single mode of the LLL is highly occupied~\cite{Fletcher2021Jun} or for other experimentally relevant initial states that are restricted to the LLL. In order to numerically compute the correlation functions of the high-temperature state, we sample product states from the respective Hilbert space sector with fixed particle density of $\nu = \tilde{\mathcal{N}}/N_y = 1/2$ and dipole moment $\tilde{\mathcal{P}}=0$ (modulo $N_y$), corresponding to the largest sector. This sector contains one of the two equivalent charge density wave states. In fact, in the thin-torus limit, $\sigma\sim 0$, these charge density wave states correspond to the bosonic FQH ground states at $\nu = 1/2$~\cite{Bergholtz2005}. To set the matrix elements $V^\text{lat}_{k,p}$, we fix the ratio $\alpha=Q/N_y=2$ and vary the effective length scale $\sigma$ by tuning $t_y/t_x$.  We then compute the autocorrelation function of the density 
$C^0(t)=\langle\Delta n_0(t) \Delta n_0(0)\rangle$, where $\Delta n_0= b^\dagger_0b_0- \langle b^\dagger_0b_0\rangle{}_{\rho_0}$ and $\langle...\rangle{}_{\rho_0}$ is the average over the sector; see Fig.~\ref{fig:3}. The data is obtained for $N_y=22$; finite size effects are discussed in App.~\ref{app:scaling}. 
After some state-dependent transient dynamics, the autocorrelation function approaches a powerlaw for late times in the thick torus limit of  large $\sigma\gtrsim 1$ with exponent of $\sim 1/4$. As will be shown later, this result is consistent with the late-time behavior of a dipole-moment conserving system. Decreasing $\sigma$, the dynamics remains stuck on the accessible time scales indicating a prethermal-regime of strong Hilbert-space fragmentation~\cite{Sala2020}.

The late-time relaxation dynamics is solely determined by the conservation laws and follows a hydrodynamics description. For a dipole-conserving systems, conventional diffusive hydrodynamics does not hold because of the restricted mobility of the system. In particular, the mobile objects are dipoles while single particles cannot move~\cite{Feldermeier2020}. From Noether's theorem for particle number conservation one has $\partial_t\rho=-\partial_x J$, where $\rho$ is the coarse grained density and $J$ is the particle current on the effective one-dimensional lattice. Since the mobile objects are dipoles, a finite charge current $J$ is obtained when dipoles are moving. The particle current $J$ is related to the dipole current $J_d$ as $J=-\partial_x J_d$, as dipoles can be interpreted as particle-hole composites~\cite{Zechmann23, Lake_2023, Gromov2020}.  The microscopic definition of the dipole current for our specific model can be computed from the Heisenberg evolution of the local density operator. One obtains $\frac{d}{dt} n_x(t)= \partial_x^2 J_d(x) $, where $J_d(x)\approx-\frac{i}{\hbar}\sum_{j,k} V_{j,k} j\, k\, (d_{j,k}(x)-\text{h. c.})$, $d_{j,k}(x)= b_{x}b^\dagger_{x+j}b^\dagger_{x+k}b_{x+j+k}$ are the dipole hopping operators, and we approximated the lattice derivative as $f(x+a)-f(x)=a\partial_x f(x)$.

As a next step, hydrodynamic assumptions allow us to express the dipole current in terms of density derivatives. In particular, as the dipole current is invariant under spatial inversion, the power of the spatial derivative needs to be even, leading to the lowest-symmetry allowed expansion $J_d=-D\partial_x ^2\rho $~\cite{Feldermeier2020, Burchards2022}. Thus, one obtains the hydrodynamics equation for dipole-conserving systems~\cite{Feldermeier2020, Gromov2020} 
\begin{equation}\label{h1}
   \partial_t \rho(x,t)= -D \partial_x^4 \rho(x,t).
\end{equation} 
In the presence of long-range, powerlaw decaying couplings, recently modifications of this hydrodynamic description have been derived~\cite{Morningstar2023,Gliozzi2023,Ogunnaike_2023}. 
In our case, the coefficients of the projected Hamiltonian, however, decay exponentially which leads the effective fracton hydrodynamics of Eq.~\eqref{h1} unchanged.

We now write the evolution of the local density as 
\begin{equation}\label{h5}
  \rho(x,t)= \int dx' K(x-x', t) \rho(x',0),
\end{equation}
where $K(x,t)$ is the kernel for the evolution. Writing Eq.~\eqref{h5} in momentum space turns the convolution into a product $\rho(q,t)= K(q, t) \rho(q,0)$, where $ \rho(q,t)=\int dx e^{iqx} \rho(x,t)$ and the Kernel $K(q,t)=\int dx e^{iqx} K(x,t)$. The hydrodynamic equation $ \partial_t K(q,t)= -D q^4 K(q,t)$ has the solution $K(q,t)=e^{-Dtq^4}$, and $K(x,t)=\frac{F(x^4/t)}{Dt^{1/4}}$ where the function $F(\cdot)$ can be written in terms of hypergeometric functions. By considering a long time approximation, one obtains $C^0(t) \sim \langle \rho(0,t) \rho(0,0) \rangle \sim \frac{1}{Dt^{1/4}}$.
Our numerical results for the squeezing Hamiltonian are thus consistent with the prediction of fracton hydrodynamics; see Fig.~\ref{fig:3}. 
With increasing typical interaction length scale $\sigma$, the time at which the hydrodynamic relaxation kicks in decreases. This is because the effective interaction is enhanced with increasing range $\sigma$.
%The hydrodynamical behavior should arise when $F(x^4/t)\to 1$, hence for times $t\gtrsim L^4$ with $L$ size of the system;  we expect that  the efficiency with which different sectors are connected is going to play a role; hence, as $\sigma$ decreases, the hydrodynamics time should increase, consistently with the results shown in~\ref{fig:3}.

\begin{figure}
  \centering
  \includegraphics[width=\columnwidth]{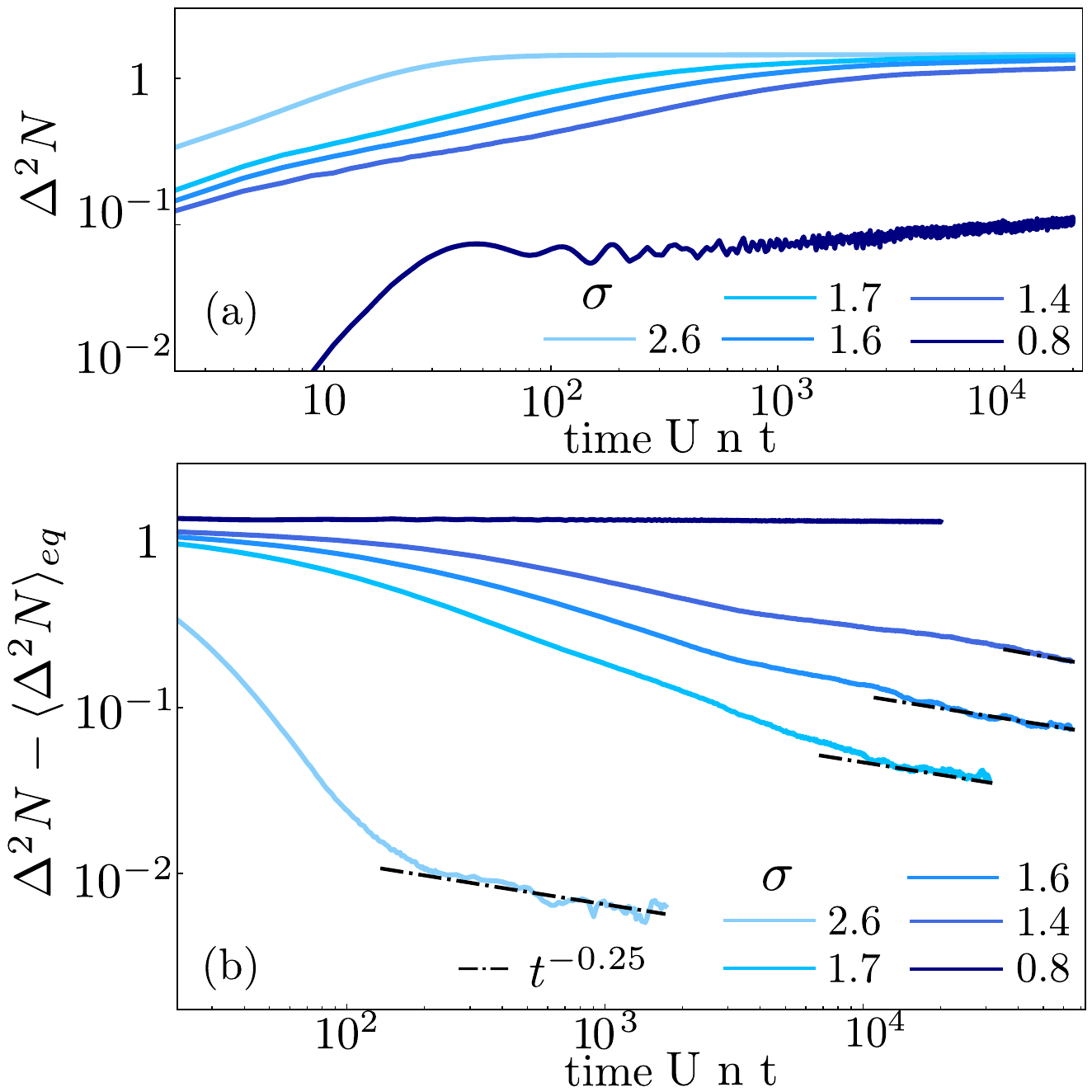}
  \caption{\textbf{Bipartite fluctuations of the particle number.} (a) Growth of the bipartite particle number fluctuations $\Delta^2 N = \sum_{i,j \in A} \langle n_i(t) n_j(t)\rangle$  following a quantum quench from product states, where the partition $A$ is chosen to be half of the system. (b) The deviations of bipartite fluctuations from the equilibrium value $\Delta^2 N-\langle\Delta^2 N\rangle_\text{eq}$ signals the anomalous sub-diffusive power law decay with exponent $1/4$ (dashed lines). 
  }
  \label{fig:4}
\end{figure}

As a next step, we study the bipartite fluctuations of the particle number in a partition $A$ of the system
\begin{align}
  \Delta^2 N&:=\sum_{i,j \in A} [\langle \tilde\rho(x_i,t) \tilde\rho(x_j,t)\rangle  ]  \nonumber\\&=\sum_{i,j \in A} \big[\langle n(i,t) n(j,t)\rangle - \langle n(i,t) \rangle \langle n(j,t)\rangle\big],
\end{align} 
following a quench from an initial product state, and $\tilde \rho(x_i,t)=n(i,t)-\langle n(i,t)\rangle$. After the quench fluctuations of the particle number build up, which follow hydrodynamic predictions~\cite{Lux2014, McCullogh2023}, as recently demonstrated experimentally for conventional diffusive hydrodynamics~\cite{wienand2023}. 
From the hydrodynamic description we obtain for the two-site, equal-time correlations~\cite{Lux2014}
$\langle \tilde\rho(x,t) \tilde\rho(x',t)\rangle-\langle \tilde \rho(x)\tilde \rho(x')\rangle_\text{eq}=\int dx_1 dx_2  K(x-x_1,t) K(x'-x_2,t)(\langle \tilde\rho(x_1,0) \tilde \rho(x_2,0)\rangle-\langle \tilde\rho(x_1) \tilde\rho(x_2)\rangle_\text{eq}) )$.
When sampling over different initial product states in the particle number $(\langle\tilde \rho(x_1,0) \tilde\rho(x_2,0)\rangle-\langle \tilde\rho(x_1) \tilde\rho(x_2)\rangle_\text{eq})\propto \delta(x_1-x_2)$. Using the Fourier transform of the kernel one finds $\int dx_1 K(x-x_1,t)K(x'-x_1, t)= K(x-x',2t)$. Then, the time dependence of the fluctuations are given by $\langle \tilde\rho(x,t) \tilde\rho(x',t)\rangle -\langle \tilde\rho(x) \tilde\rho(x')\rangle_\text{eq} \sim K(x-x',2t)$. After integration over the bipartition, we obtain for times $t \gg N_y^4/D$ at which 
$F(x^4/t)\to 1$
\begin{equation}\label{eq:13}
 \Delta^2 N - \langle \Delta^2 N \rangle_\text{eq} \sim \frac{1}{(2Dt)^{1/4}}.
\end{equation}
Thus the fluctuations approach their finite-size saturation value hydrodynamically.

We evaluate the fluctuations numerically following a quantum quench of a product initial state. The bipartite number fluctuations $\Delta^2 N$ grow with time and saturate to a finite value due to the finite system size; Fig.~\ref{fig:4} (a). We can best extract the hydrodynamic relaxation dynamics by directly computing the deviations of the fluctuations from their late-time saturation value; Fig.~\ref{fig:4} (b). We observe that similarly to the autocorrelation function, the hydrodynamic behavior in the fluctuations sets in at later times as the interaction length scale $\sigma$ is decreased. For $\sigma=1.4$ the onset time of hydrodynamics is difficult to estimate due to the very slow relaxation in the non-universal regime. For sufficiently small $\sigma\lesssim 1$, the hydrodynamic regime is not observable on the time scales and system sizes accessible in our numerical simulations;  as the dominant terms in the model are of shortest possible range giving rise to strong Hilbert space fragmentation with exponentially many disconnected Krylov subspaces~\cite{Sala2020}.  Nonetheless, due to the Gaussian decaying coupling terms, strong Hilbert space fragmentation is always formally avoided in our model for any finite interaction length scale $\sigma$, implying that in the thermodynamic limit and at late times the system will ultimately be able to thermalize.

\section{Fractional Quantum Hall Physics with Ultracold Atoms \label{Ultracold}}
Here, we explore how the dynamics is affected by considering all Landau levels and investigate the role of inter-level transitions. We particularly focus on two types of experimental platforms: rotating quantum gases (Sec. \ref{Rotating}) and ultracold atoms in optical lattices (Sec. \ref{UltracoldLattice}); and consider the full unprojected continuum and lattice models. Our analysis also clarifies the connection between the effective projected one-dimensional lattice and real space. 

The squeezing terms of the effective one-dimensional models projected to the LLL are expressed in momentum basis, see Eqs.~\eqref{eq:LLLcont} and \eqref{eq:projH}. However, both in the continuum and on the lattice, wave functions corresponding to different momenta $q$ in the LLL are the same apart for a translation in the $x$-direction (which we discuss for the lattice in App.~\ref{dcrealspace}, while for the continuum this can be directly read off from the wave function). As a consequence, we can equally interpret the projected Hamiltonians to describe squeezing motion in $x$-direction of real space and, hence, the dipole moment is also conserved in real space along the $x$-direction $\mathcal P = \sum_{x,y} x n_{x,y}$.  The aim of this section is to analyze %when We will now further elaborate on this point and analyze the full unprojected dynamics. This will allow us further to determine 
the minimal requirements for observing squeezing dynamics in real space for both rotating quantum gases and ultracold atoms in optical lattices, when the full unprojected dynamics is considered. The initial states of the dynamics here considered are states where all the particles are initialized in a finite energy density state in the LLL, which will give rise to a similar relaxation dynamics to the ones considered in the previous section.

 \subsection{Rotating quantum gases in the continuum}\label{Rotating}

Using the analogy between the Lorentz force and the Coriolis force, synthetic gauge fields can be realized by rotating quantum gases. 
Consider a Bose-Einstein condensate that is confined in a three-dimensional harmonic trap with weak isotropic in-plane confinement $\omega_x=\omega_y=\omega$ and strong vertical confinement $\omega_z\gg \omega$. In addition, the trap rotates around the vertical axis with an angular frequency $\Omega$. The single particle Hamiltonian is thus given by $H=\frac{1}{2m}\big[p^2_x+p^2_y+m^2 \omega^2(x^2+y^2)\big]-\Omega L_z$,
where $L_z$ is the axial angular momentum operator. The rotational term $\Omega L_z$ is mathematically equivalent to an applied magnetic field in the vertical direction that in the symmetric gauge has the form $\bold A= m \Omega(-y, x)$: $H=\frac{1}{2m}\big[(\bold p-\bold A)^2 +m^2(\omega^2-\Omega^2)(x^2+y^2)\big]$.
When a small anisotropy $\epsilon$ is introduced in the in-plane trapping potential, $\omega_x = \sqrt{1+\epsilon}\,\omega$ and $\omega_y = \sqrt{1-\epsilon}\,\omega$, the single particle Hamiltonian becomes
\begin{equation}
     H=\frac{1}{2m}\big[(\bold p-\bold A)^2 +m^2((1+\epsilon)\omega^2-\Omega^2 )x^2+m^2((1-\epsilon)\omega^2-\Omega^2 )y^2\big].
\end{equation}
The anisotropy in the trapping potential can be used to prepare a state in the LLL~\cite{Fletcher2021Jun,Mukherjee2022Jan,Crepel2023,Yao2023}. This can be seen by considering the regime of $\Omega=\omega$, and small value of the anisotropy, for which the evolution is equivalent to a squeezing operation. In that limit, the single-particle Hamiltonian is given  by~\cite{Fletcher2021Jun}
\begin{equation}
    H= 2\hbar \omega( \alpha^\dagger \alpha+\frac{1}{2})+\frac{ \hbar\epsilon \omega }{4}( \beta^\dagger \beta^\dagger+\beta\beta),
\end{equation}
where $\alpha=\frac{1}{2}\sqrt{\frac{m\omega}{\hbar}}(x-\frac{p_y}{m\omega}+i(y+\frac{p_x}{m\omega}))$  and $\beta=\frac 1 2 \sqrt{\frac{m\omega}{\hbar}}(x+\frac{p_y}{m\omega}-i(y-\frac{p_x}{m\omega}))$ are the lowering   operators for the single particle excitations in cyclotron and guiding center coordinates, respectively. The evolution thus determines a geometric squeezing of the guiding center phase space distribution, and effectively drives the condensate into the LLL~\cite{Fletcher2021Jun,Crepel2023}.
Once the condensate is squeezed into the LLL, which has been recently achieved experimentally~\cite{Fletcher2021Jun}, the anisotropy can be turned off, the system starts evolving due to contact interactions $U$ inside the LLL. The interaction-induced dynamics remains confined to the LLL in the limit of large separations between the bands, $Un \ll \omega_c$, where the cyclotron frequency $\omega_c=2\Omega$, and is thus governed by the effective slow relaxation dynamics discussed in Sec.~\ref{dynamicsLLL}.

In order to gain some insight into the dynamics of particles prepared in the LLL, we can consider a minimal model of two particles in a magnetic field, $H=H_1+H_2=\frac{1}{2}\big[(\bold p_1-\bold a_1)^2+(\bold p_2-\bold a_2)^2\big]$ where $\bold a_{1,2}=(0, x_{1,2})$  in the Landau gauge. The Hamiltonian is separable in center of mass $\bold P, \bold R$ and relative $\bold p, \bold r$ degrees of freedom, $H=\frac{1}{2} \big [\frac{1}{2}(\bold P -2 \tilde{\bold{ A}})^2+ 2(\bold p -\frac 1 2 \bold a)^2\big]$, where $\tilde {\bold A}=(0, X)$ and $\bold a =(0,x)$, with spectrum $E_{\gamma_1,\gamma_2}=\hbar\omega_c (\gamma_1 +\gamma_2+1)$. 

It follows that, when the initial state is fully contained in the LLL, both the center-of-mass wave function and relative distance wave function will be eigenstates in the LLL. For density-density interactions only the relative distance wave function evolves, while the center-of-mass wave function is still an eigenstate of the Hamiltonian. This implies that, even in the presence of interband transitions, as long as the initial center of mass component of the wavefunction lives in a single LL, the real space dipole in the $x$-direction is conserved by the dynamics. Thus, both average center of mass 
and its higher-moments are constant, and the dipole moment in real space is preserved to all orders. This reasoning can be directly generalized to more particles as well. 
\begin{figure}
    \centering
    \includegraphics[width=\columnwidth]{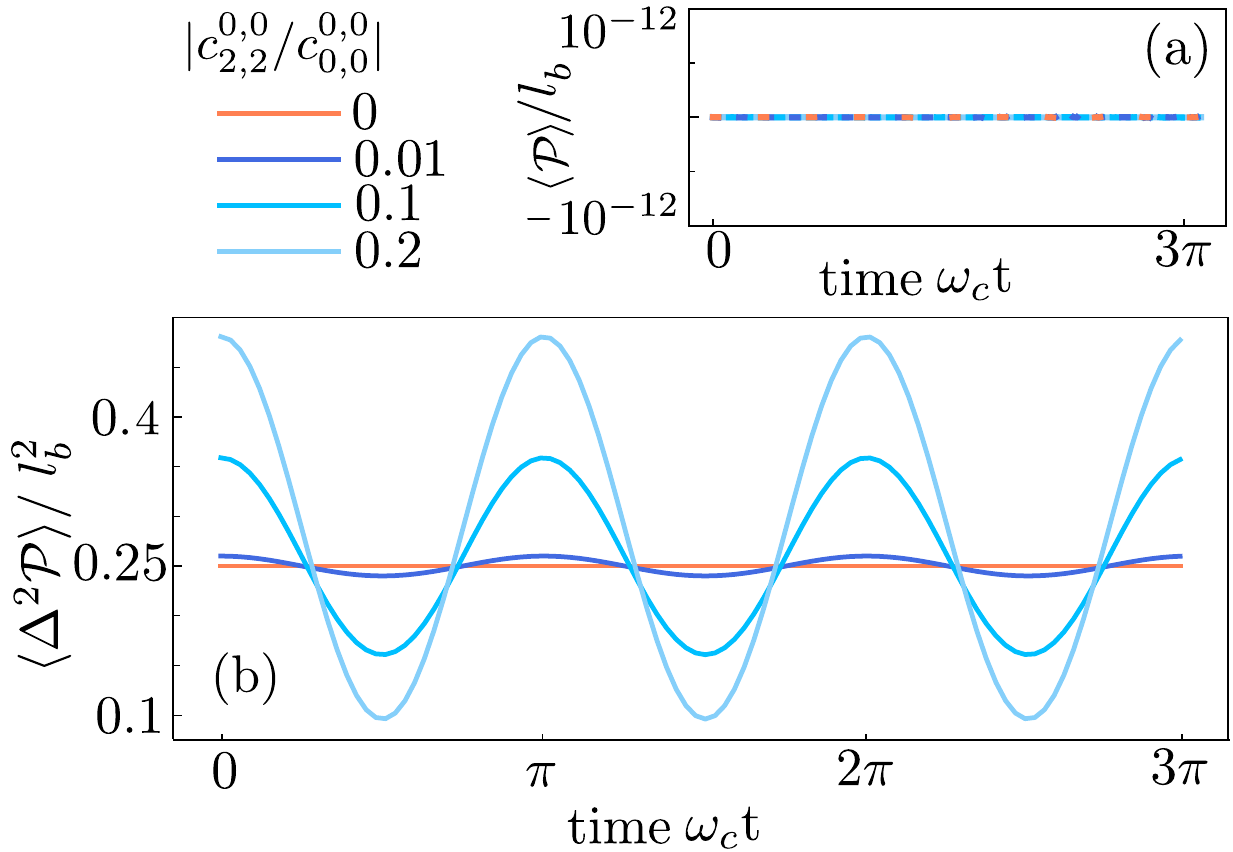}
    \caption{\textbf{Dipole-moment dynamics in the continuum model}. Evolution of the  (a) average of the real-space center of mass (or dipole moment) $\langle \mathcal P\rangle$ and (b) its fluctuations $\langle \Delta \mathcal P^2\rangle=\langle \mathcal P^2\rangle-\langle \mathcal P\rangle^2$  for an initial state $\Psi_0(\bold R, \bold r)= c^{0,0}_{0,0} \psi^0_{0}(\bold R) \phi^{0}_0  (\bold r)+ c^{0,0}_{2,2} \psi^0_{2}(\bold R) \phi^{0}_2  (\bold r)$, for different ratios of the coefficients $|c^{0,0}_{2,2}/c^{0,0}_{0,0}|$. }
    \label{figrot}
\end{figure}

When the center-of-mass wave function of the initial state is not an eigenstate, i.e., when the initial state of the center of mass is not confined in the LLL, higher moments of the dipole moment will not be conserved in general. We consider an initial state that is a superposition of different Landau levels both in the center of mass and relative coordinate $\Psi(\bold R,\bold r )=\sum_{\gamma \beta, n , m} c_{\gamma,\beta}^{n,m} \psi_\gamma^{n}(\bold R) \phi_{\beta}^m  (\bold r)$, where $\psi(\bold R)$ and $\phi( \bold r)$ are eigenstates of the non-interacting problem for the center of mass and relative coordinate, respectively. Specifically they are given by
\begin{align}
    \psi_{\gamma}^{n}(\bold R)=\frac{1}{\sqrt{{\pi}\gamma!2^\gamma L_y}}H_\gamma\bigg(\frac{X-X_n}{L_B}\bigg)e^{-\frac{1}{2}(\frac{X-X_n}{ L_b})^2}e^{iK_n Y}
\end{align}
where $K_n=2\pi n/L_y$, $X_n=L_b^2 K_n$, and the magnetic length for the center of mass coordinate $L_b=\frac{l_b}{\sqrt{2}}$, as well as,
\begin{align}
    \phi^{m}_{\beta}(\bold r)=\frac{1}{\sqrt{{\pi}\beta!2^\beta L_y}}H_\beta\bigg(\frac{x-x_m}{\ell_b}\bigg)e^{-\frac{1}{2}(\frac{x-x_m}{ \ell_b})^2}e^{ik_my}
\end{align}
where $k_m=2\pi m/L_y$, $x_\beta=\ell_b^2 k_m$, and the magnetic length for the relative coordinate $\ell_b=l_b\sqrt{2}$. By using the properties of the Hermite polynomials $H_n$ we  determine the evolution of the moments of the center-of-mass in real-space in the $x$-direction (i.e., dipole moment in the $x$-direction in real space) $\mathcal{P} = X$ and find that for general initial states the center of mass is not conserved.

As a concrete example, we consider an initial state where the two particles have momenta $k_y=0$ and that is invariant under inversion symmetry in the spatial coordinate. Since the two particles are bosons the wave function $\Psi(\bold R,\bold r )=\sum_{\alpha, \beta, n , m} c_{\gamma,\beta}^{n,m} \psi_\gamma^{n}(\bold R) \phi_{\beta}^m  (\bold r)$, where $\psi(\bold R)$ and $\phi( \bold r)$ must be invariant under exchange of their coordinates $\Psi (\bold R,- \bold r)=\Psi (\bold R, \bold r)$, which implies $m$ must be even. Considering the state to be invariant under inversion symmetry, $\Psi (-\bold R, -\bold r)=\Psi (\bold R, \bold r)$, $n$ is even as well. Under these assumptions, the average position of the center of mass does not evolve. For concreteness, we show the evolution of the initial state $\Psi_0(\bold R, \bold r)= c^{0,0}_{0,0} \psi^0_{0}(\bold R) \phi^{0}_0  (\bold r)+ c^{0,0}_{2,2} \psi^0_{2}(\bold R) \phi^{0}_2  (\bold r)$, for different ratios of $|c^{0,0}_{2,2}/c^{0,0}_{0,0}|$ in Fig~\ref{figrot}. %As expected from Eq.~\eqref{P2}, 
The fluctuations of the center of mass oscillate with frequency $2 \omega_c$ when multiple Landau levels are occupied. When the initial state is fully contained in the LLL, $c^{0,0}_{2,2}=0$,  the fluctuations of the dipole-moment $\langle \Delta \mathcal{P}^2\rangle$ (as well as all higher moments) are constant, as argued before. When creating such an initial state in the LLL with a finite particle density, the subsequent relaxation dynamics in real space will be governed by fracton hydrodynamics; see discussion in the previous section.

\subsection{Synthetic gauge fields for ultracold atoms in optical lattices}\label{UltracoldLattice}
Charge neutral atoms in optical lattices with artificial gauge fields are effectively described by the interacting Harper-Hofstadter model, Eq.~\eqref{eq:1}, and geometric conditions on the lattice can be defined such as the non-interacting bands are flat. 
\begin{figure}
  \centering
  \includegraphics[width=0.97\columnwidth]{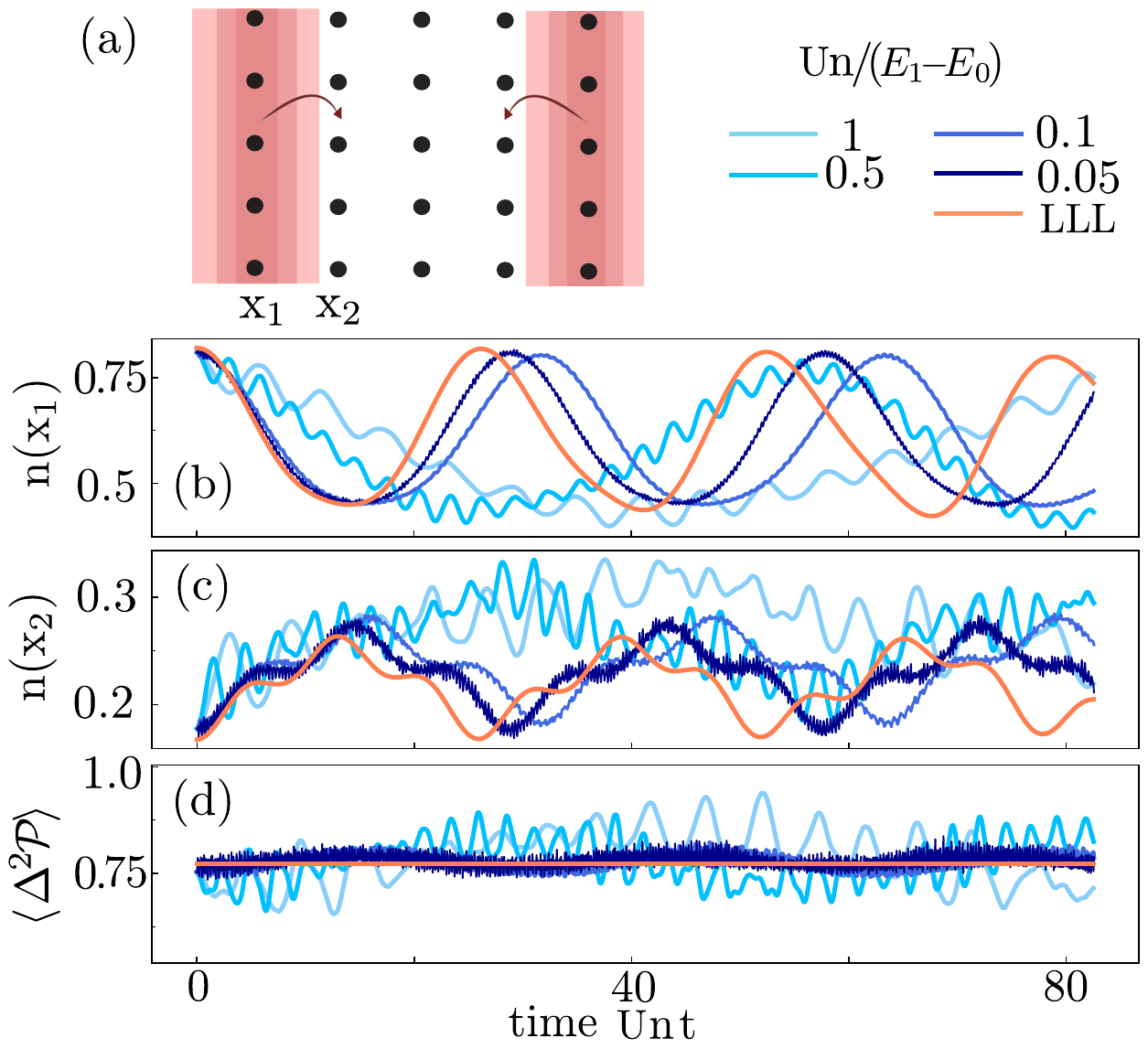}
  \caption{\textbf{Emergent dipole-moment conserving dynamics.} (a) Schematic illustration of the two-particle initial state in the 2D lattice. (b) and (c) Evolution of the column density centered at $x_1$ and $x_2$ respectively, as indicated in (a). (d) Fluctuations of the center of mass indicate an emergent dipole-moment conservation as the strength of the interaction is decreased. 
  The fluctuations of the dipole moment of the initial state are finite, as it is not an eigenstate of the dipole operator.}
  \label{fig:2}
\end{figure}
We now investigate the full dynamics of this model and initialize the system at time $t=0$ with particles occupying orbitals of the lowest band of the non-interacting problem $H_0$. Experimentally, such a state can be obtained by preparing plane-wave states in the $y$-direction, corresponding to an eigenstate of the non-interacting Hofstadter model with ${t_x}/{t_y}\to 0$, followed by a slow increase of ${t_x}/{t_y}$; \textit{c.f.} Ref.~\cite{Leonard2023} where a similar approach has been considered. At time $t=0$ the on-site interaction energy is turned on  and the particles start to evolve.  

In order to gain insights into the effective dynamics, we simulate a minimal model for two particles and $(N_x,N_y) = (5,5)$ using exact diagonalization; see Fig~\ref{fig:2}(a).  
The column density centered around the coordinate $x_i$ exhibits an oscillatory dynamics associated with the average conservation of the center of mass; see Fig.~\ref{fig:2}(b,c).
The oscillation frequency is given by the effective matrix elements of the local interaction $\sim U n$ which we use to rescale time.
We compare the dynamics of the unprojected Hamiltonian Eq.~\eqref{eq:1} with the effective one projected onto the LLL, Eq.~\eqref{eq:projH}. The lower the interaction, the closer the effective dynamics and the full dynamics become.  This is because interband transitions are suppressed with decreasing interactions; as relevant energy scales  the matrix elements of the local interaction $\sim U n$ have to be compared with the cyclotron frequency, i.e., with the energy difference of the LLL to the first LL. We observe that as $U$ decreases, the period of the evolution decreases as well, meaning that the rescaled energy is increasing. This is because the second order corrections to the eigenvalues of the interacting Hamiltonian due to interband processes on the ground state is negative, leading to a larger oscillation period. 

Contrary to the continuum case, the Hamiltonian is not separable in center of mass and relative coordinate degrees of freedom. Thus, although the states exhibit an average conservation of the center of mass along the $x$-direction, $\mathcal{P} =\sum_{x,y} x \, n_{x,y}$, when their density profile is invariant under inversion symmetry (not shown), the fluctuations $\langle\Delta^2\mathcal{P}\rangle=\langle \mathcal{P}^2\rangle-\langle\mathcal{P}\rangle^2$ will not be constant in the presence of interband transitions (even when initially prepared in the LLL). On the other hand, when the dynamics is projected in the LLL,  the value of the fluctuations  (as well as higher momenta of the center of mass) is preserved. 
This is because similarly to the continuum the shape of the wave functions in the LLL remain invariant for different momenta and only their wave function center shifts along the $x$-direction, see App.~\ref{dcrealspace}. Hence, a squeezing motion of the column density is expected in real space provided the dynamics is confined to the LLL. 
This formally follows also from the fact that the center of mass along the $x$-direction $\mathcal P$ projected onto the LLL, commutes  with the projected interactions, see App.~\ref{dcrealspace}.
By contrast, for interlevel processes $\mathcal{P}$ is not conserved in real space, despite the total momentum $\sum k_y n_{k_y}$ being conserved, where the occupation of momentum $k_y$, $n_{k_y}=\sum_\gamma n_{k_y,\gamma}$, is obtained by summing over all Landau levels $\gamma$. Eigenfunctions with the same $k_y$ but on different Landau levels thus have different expectation values of the dipole-moment $\mathcal{P}$ evaluated in real space. We compute numerically the fluctuations of the dipole moment $\langle\Delta^2\mathcal{P}\rangle=\langle \mathcal{P}^2\rangle-\langle\mathcal{P}\rangle^2 $ and find that they are reduced with decreasing interactions as the dynamics becomes increasingly confined to the LLL, see Fig.~\ref{fig:2}(d), consistent with the  discussion above. In summary, on the lattice the real-space dipole moment is only conserved when the dynamics is limited to one LL, since interband processes do not preserve dipole moment (center of mass) in real space. Thus, for observables integrated over the y-direction but local in the $x$-direction, a subdiffusive relaxation dynamics will arise only in the limit of small interactions compared to the difference of energies between different bands.

\section{Discussion and Outlook}\label{outlook} 

Recent experiments with ultracold atoms have realized models with synthetic gauge fields. In these systems,  non-equilibrium quantum dynamics are most naturally accessible. For this reason, it is crucial to understand relaxation processes for states with partially filled Landau levels. 

In contrast with the common anticipation that interacting quantum many-body systems relax diffusively, we showed that this is not the case for finite energy density states in the lowest Landau level. Instead, the relaxation dynamics of finite energy density states confined within a Landau level effectively occurs in a one-dimensional lattice and is constrained by dipole conservation in the effective lattice. Hence, the far-from-equilbrium dynamics in the LLL are governed by fracton hydrodynamics, characterized by a slow subdiffusive relaxation of density excitations and their fluctuations. The properties of the relaxation are  tunable by the effective thickness of the torus. The equivalence of dipole conservation in the LLL and dipole conservation in real space along the $x$-direction allows for a direct measurement of the unconventional relaxation dynamics,  both in rotating quantum gases and in interacting Harper-Hofstadter models with certain geometric constraints. 

For future work, it will be interesting to analyze the short time dynamics of specific initial states in more detail. Furthermore, one could study the consequences of non-flat bands in the interacting Harper-Hofstadter model, that arise for example with open boundary conditions, on the effective relaxation dynamics. Moreover,  
excitations of the FQH system could be studied in the low-temperature regime and related to the different dynamical regimes of fracton excitations~\cite{zechmann2023dynamical, boesl2023}.  It would be also exciting to explore relaxation processes in the vortex dynamics of rotating quantum gases~\cite{Fletcher2021Jun}. Previous works have already interpreted vortices as fractons and discussed Hilbert-space fragmentation phenomena and fractonic dynamics in vortex systems~\cite{Doshi2021}.

\textbf{Data and Code availability.---}Data analysis and simulation codes are available on Zenodo upon reasonable request~\cite{zenodo}.

\textbf{Note added.---}While finalizing our manuscript, we became aware of an experimental work which studies the interaction induced dynamics of a condensate initially prepared at zero momentum in the LLL~\cite{Yao2024}. This experiment observes the squeezing dynamics for a specific initial state. 

\textbf{Acknowledgements.---}We thank S. Chi, R. Fletcher, N. Goldman, P. Ledwith, R. Yao, and especially M. Zwierlein for many insightful discussions. We acknowledge support from the Deutsche Forschungsgemeinschaft (DFG, German Research Foundation) under Germany's Excellence Strategy--EXC--2111--390814868, DFG grants No. KN1254/1-2, KN1254/2-1, and TRR 360 - 492547816, FOR 5522 (project-id 499180199) and from the European Research Council (ERC) under the European Unions Horizon 2020 research and innovation programme (Grant Agreements No. 771537 and 851161), as well as the Munich Quantum Valley, which is supported by the Bavarian state government with funds from the Hightech Agenda Bayern Plus. A.S. acknowledges support by the National Science Foundation under Grant No. DMR-2029401.

\appendix
\begin{figure*}
  \centering
  \includegraphics[width=0.96\textwidth]{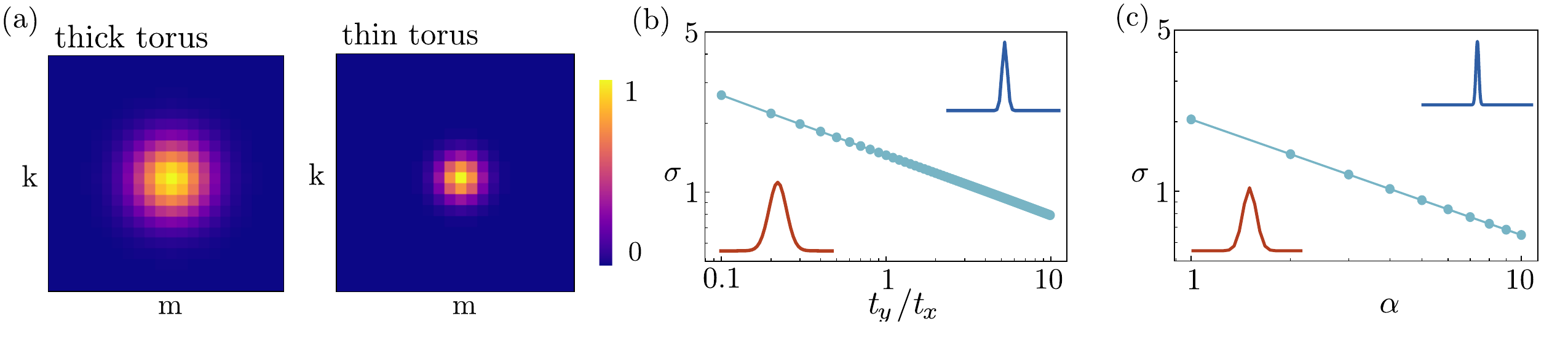}
  \caption{\textbf{Analysis of the potential.} (a) $V_{k,m}/V_{0,0}$ for the thick and thin torus limit, respectively. (b) Width of the potential $\sigma$ as a function of the hopping $t_y/t_x$ for $\alpha=2$ and $N_y=22$. (c) Width $\sigma$ as a function of $\alpha$ for $t_y/t_x=1$ and $N_y=22$. The error bar of the fit is smaller than the size of the markers. Insets: wave functions in the $x$-direction in the thick and thin torus limit, respectively.}
  \label{fig:sup1}
\end{figure*}

\section{Projection onto the lowest Landau level in the flat-band lattice model}\label{ProjectionLLL}

On the lattice, we consider a generic density-density interaction of the form ${V}=\sum_{\bold{q}} V(\bold{q})\rho(\bold q)\rho(-\bold q)$,
where $\rho(\bold q)=\sum_{x,y}e^{i(\frac{2\pi}{Q} q_x x+\frac{2\pi}{N_y} q_y y)} \rho(x,y)$ is the Fourier transform of the density operator, and $\bold q =(q_x,q_y)$, $q_x \in 0,...,Q-1$, $q_y \in 0,....,N_y-1$ are in the original Brillouin zone of the lattice. The projection of the density operator onto lowest band gives~\cite{Scaffidi2014}\begin{equation}
    \tilde{\rho}_0 (\bold q)=P_0\rho(\bold q) P_0=\rho_0(\bold q) d_n (\bold q),
  \end{equation}
where $\rho_0(\bold q)$ fulfils the Girvin-MacDonald-Platzman (GMP) algebra $[\rho_0(\bold q_1),\rho_0(\bold q_2)]=2i \sin (-\frac{1}{2}\frac{2\pi}{N_y}C \bold q_1 \times\bold q_2)\rho_0(\bold q_1+\bold q_2)$~\cite{Girvin1986} and acts on a state of the LLL as
  $\rho_0(\bold q)|{K_y}\rangle=e^{-\frac{i}{2}\frac{2 \pi}{N_y}Cq_xq_ye+i\frac{2\pi}{N_y}CK_yq_x}|{K_y+q_y}\rangle$.  
We also introduced the form factor 
  $d_0(\bold q)=e^{-i\frac{1}{2}\frac{2\pi}{N_y}C q_x q_y}\sum_x e^{i2\pi\frac{x}{Q}q_x}u^*_{0,0}(x)u_{0,0}(x+C\alpha q_y)$.
Projecting the Hamiltonian onto the LLL we thus obtain
\begin{equation}
  P_0HP_0=\sum_{\bold q} \tilde{V}(\bold q) \rho_0(\bold q)\rho_0(-\bold q)
\end{equation}
where $\tilde{V}(\bold q)=V(\bold q)| d_0(\bold q)|^2$. For contact interactions, $V(\bold q)={U}/N_xN_y$.
We can rewrite this expression as
\begin{multline}\label{eq:9}
  P_0HP_0=\sum_{q=0}^{N_y-1} \sum_{p=0}^{\frac{N_y}{2}-1}\sum_{k=p+1}^{\frac{N_y}{2}} \frac{V_{k,p} b^\dagger_{q}b^\dagger_{q+k+p}b_{q+k}b_{q+p}}{(1+\delta_{p,0})(1+\delta_{k,\frac{N_y}{2}})} + \text{h.c.}
\end{multline}
where we introduced the creation (annihilation) operators for the bosons in the lowest band $b^\dagger_k(b_k)$ and $k,p,q$ are momenta in the $y$-direction. The projected Hamiltonian effectively conserves the dipole moment on that effective one-dimensional lattice.
Explicitly, we obtain\begin{align}
  & V_{k,p}= \frac{U}{A}\sum_{q_x} |d_0(q_x, k)|^2e^{-i\frac{2\pi}{N_y}Cq_xp}+\nonumber\\&|d_0(q_x, -k)|^2e^{i\frac{2\pi}{N_y}Cq_xp}+|d_0(q_x, p)|^2e^{-i\frac{2\pi}{N_y}Cq_xk}+\nonumber\\&|d_0(q_x, -p)|^2e^{i\frac{2\pi}{N_y}Cq_xk}.\end{align}
 Thus, when projecting onto the lowest band we obtain an effective dipole preserving potential on a one dimensional lattice. In Eq.~\eqref{eq:9} we report the potential for hard-core bosons in the Lowest band, as considered in the main text.

We observe that $V_{k,p}$ depends on the non-interacting Hamiltonian via the form factor. For contact interactions and geometries of the lattice $P=1$, $Q=\alpha N_y$, the projected interactions decay approximately as $\sim e^{-\frac{k^2+p^2}{2\sigma^2}}$, 
and the parameter $\sigma$ can be numerically estimated, see Fig.~\ref{fig:sup1}. The dipole squeezing moves are suppressed by the Gaussian shaped potential as one increases the distance $|p-k|$ of the particles and the distance of the hoppings $p$. 
When increasing the overlap between eigenstates of the non-interacting Hamiltonian, the width $\sigma$ increases as well. The width of the eigenfunctions can be changed by changing either $\alpha =Q/N_y$ or the hopping ratio $t_y/t_x$, see Fig~\ref{fig:sup1} (b,c). We have used the latter to change the relaxation properties of the system from a hydrodynamic to a prethermal regime. The dynamics can be similarly tuned by modifying the ratio $\alpha$ of the physical dimensions of the system. Thus, changing the thickness of the torus or the ratio between the hopping matrix elements in the $x$ and $y$ direction is equivalent.\\

\section{Generalization to different geometries and non-flat bands}\label{NonFlatbands}
The analysis  in the main text focuses on the flat band condition, obtained directly for rotating quantum gases or  for specific choices of lattice geometries in the Harper-Hofstadter model. Here, we generalize the considerations for lattice systems to non-flat bands. Non flat-bands arise in a lattice when the condition $N_x=Q=\alpha N_y$, is not satisfied and when the boundary conditions are not met. However, for cases where  the width of the band $\Delta$ is small compared to the other energy scales, $\Delta\ll U\ll \hbar \omega_c$, our expectation is that this will not modify the thermalization properties in the LLL.

We first consider $N_x=Q\not=\alpha N_y$ and a torus geometry with periodic boundary conditions in the $x$- and $y$-direction. The non-interacting Hamiltonian $H_0$ is invariant under magnetic translations, see Eq.~\eqref{eq:2-3}, and $[T_x^Q,T_y]=0$. Imposing periodic boundary conditions both in the $x$- and $y$-direction, we obtain $Q$ bands with $N_y$ eigenstates characterized by the quantum numbers $k_y$, $\psi_{k_y,n}(x,y)=e^{-i\frac{2\pi}{N_y}k_yy} u_{k_y, n }(x)$ where $k_y \in [0,...N_y-1]$. However, in this case the set of Hofstadter equations~\eqref{hofstd} do not map onto themselves when changing $k_y$. As a consequence the bands are not flat. In this case, we can still project the potential in the lowest band following a procedure similar to  before. In particular, we find that the projected density operator \begin{equation}
  \tilde{\rho}(\bold{p})=P_0\sum_{\bold r } e^{i \bold p \bold r } \rho(\bold r)P_0=\sum_{k_y}d(k_y,\bold p)b^\dagger_{k_y}b_{k_y+p_y},
\end{equation}
where $d(k_y,\bold p)=\sum_x e^{i p_x x} u_{k_y,n}(x)u_{k_y+p_y,n}(x)$, which coincides with a distance in the effective lattice apart for a phase factor dependent on $k_y$. Moreover,  $b^\dagger_{k_y}$($b_{k_y})$ are, as usual, creation (annihilation) operators in the lowest band. Then the projected potential is 
\begin{equation}
  P_0VP_0=\sum_{p,k,k'} \tilde{V}_{k,k',p}b^\dagger_{k}b_{k+p}b^\dagger_{k'}b_{k'-p},
\end{equation}
which is dipole preserving in the effective one-dimensional lattice. However, the single-particle band has a finite width, and this needs to be considered in the effective dynamics, leading to a potential that further reduces the relaxation dynamics as long as the its strength is comparable with the effective interaction. 

For $N_x\not =Q=\alpha N_y$, and $N_x=\beta Q$ the eigenstates of the non interacting problem $|{ k_x,k_y,n}\rangle=e^{i\frac{2\pi k_x}{\beta}\frac{x}{Q}}e^{i\frac{2\pi}{N_y}k_y y } u_{k_x,k_y,n}(x)$ are defined also by the quantum number $k_x \in 0,...,\beta-1 $, as well as the usual $k_y \in 0,...,N_y-1$  and $n$ the  band index. The set of Hofstadter equations does not map onto itself when $k_y$ and $k_x$ are changed. As before, we can project the potential onto the lowest band and obtain the dynamics in the effective 2D lattice $\beta\times N_y$. The projected density operator is given by \begin{multline}
  \tilde{\rho}(\bold{p})=\sum_{x,k_y, k_x, k_x'} e^{i \frac{2\pi p_x x}{N_x}} e^{i \frac{2\pi k_x'}{\beta}\frac{x}{Q}}\\ e^{i \frac{2\pi k_x}{\beta}\frac{x}{Q}} u_{k_x,k_y}(x)u_{k_x',k_y+p_y}(x)b^\dagger_{k_x,k_y}b_{k_x',k_y+p_y}.
\end{multline}
Projecting the density-density contact interaction onto the lowest band, we obtain the effective potential
\begin{multline}
\sum_{\bold p,k_y,k_y', k_{1,x}, k_{1,x}', k_{2,x}, k_{2,x}'} \tilde{V}(\bold p,k_y,k_y', k_{1,x}, k_{1,x}', k_{2,x}, k_{2,x}')\\b^\dagger_{k_y, k_{1,x}}b_{k_y+p_y,k_{1,x}'}b^\dagger_{k_y',k_{2,x}}b_{k_y'-p,k_{2,x}},
\end{multline}
that is dipole preserving along the $y$-direction of the effective 2-dimensional lattice. After integrating the column density, however, the dynamics becomes dipole-conserving again.

Next, we relax the requirement of periodic boundary conditions. We first discuss cylindrical boundary conditions for $N_x=Q=\alpha N_y$; other generalizations follows from the considerations above. When open boundary conditions in the $x$-direction are considered (cylinder geometry), the bands will not be flat anymore, since the set of Hofstadter equations do not map one on the other. Similarly to the above, a projection of the potential to the lowest band still gives a dipole preserving potential. The effective one dimensional lattice does not have periodic but open boundary conditions. In the large systems limit, this does not influence the relaxation behavior. For open boundary conditions in both directions analytical solutions cannot be obtained, however, we expect that for sufficiently large systems effective dipole-conserving dynamics will emerge. 
We emphasize that these considerations are only required for lattice systems. Rotating quantum gases in the continuum automatically realize flat bands and  thus more directly realize the emergent fracton hydrodynamics.

\begin{figure}
  \centering
\includegraphics[width=\columnwidth]{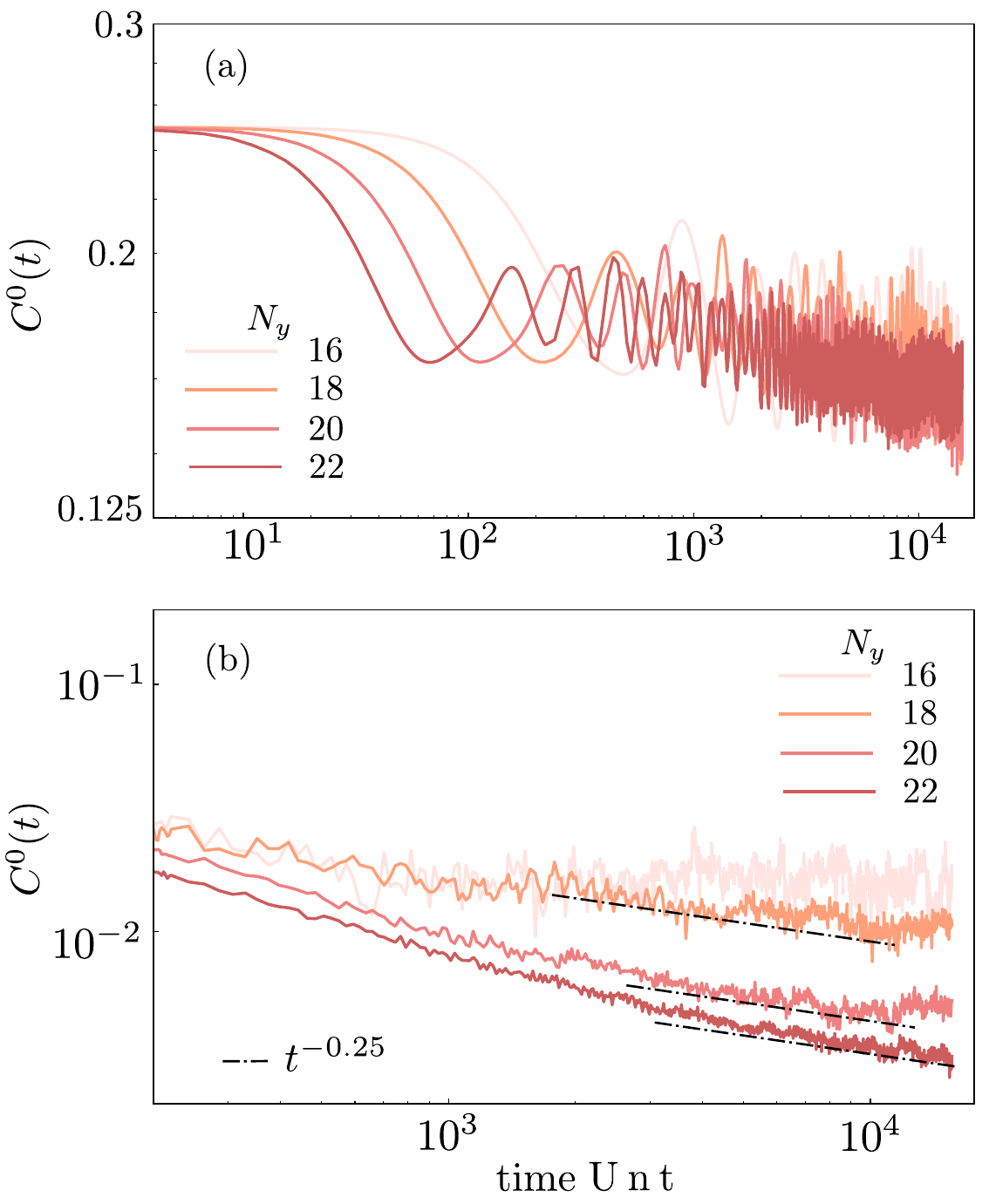}
  \caption{\textbf{Density-density autocorrelation function.} Evolution of the density-density autocorrelation function $C^0(t)=\langle\Delta n_0(t) \Delta n_0(0)\rangle$ for different values of $N_y$, $\alpha=2$ and for (a) $\sigma=0.8$ and (b) $\sigma=1.7$, where $\Delta n_0= n_0- \langle n_0\rangle{}_{\rho_0}$ and $\langle...\rangle{}_{\rho_0}$ is the average over the chosen sector of the density operator. }
  \label{fig:sup2}
\end{figure}

\section{Results for different systems sizes \label{app:scaling}} 

We evaluate the density-density autocorrelation function for different values of $N_y$ in Fig.~\ref{fig:sup2}. In the thin torus limit (a), the system does not immediately relax but instead enters a prethermal regime. As exponentially small longer-range terms are inevitably present in the Hamiltonian, the system will  ultimately relax in the thermodynamic limit. This will occur only on exponentially long scales. By contrast, in the thick torus limit (b), subdiffusive hydrodynamic relaxation is evident for sufficiently large systems. 

\section{Dipole conservation in real space }\label{dcrealspace}
As remarked in the main text, dipole conservation in the LLL can be connected to dipole conservation in real space along the $x$-direction $\mathcal P = \sum_{x,y} x n_{x,y}$, when both the dynamics and the dipole moment are projected in the LLL.  The present analysis focuses on the lattice case, since dipole conservation  in real space for the continuum model follows  from the separability of center of mass and relative coordinate degrees of freedom;  here we consider the two particle model, but this analysis can be easily generalized to arbitrary number of particles. We define $\Psi^{0,0}_{k,q}(\bold r_1, \bold r_2)= \frac{1}{\sqrt{2}}[e^{ik y_1} u_{k,0}(x_1)e^{iq y_2} u_{q,0}(x_1) + x_1 \leftrightarrow x_2 \, , \,y_1 \leftrightarrow y_2]$ the (symmetrized) wavefunction for two bosons in a product state in the LLL. The expectation value of the dipole operator in the $x$-direction,  $ \langle  \Psi^{0,0}_{k,q } \, \mathcal{P}\, \Psi^{0,0}_{p,l}\rangle=\delta_{\{k,q\},\{p,l\}}\sum_{x_1,x_2} (x_1+x_2-\alpha C(k-q) ) |u_{0,0}(x_1)|^2|u_{0,0}(x_2)|^2 $,  is the same for every wavefunction with the same dipole  moment in the LLL,  $\langle  \Psi^{0,0}_{k,q } \, \mathcal{P}\, \Psi^{0,0}_{k,q}\rangle= \langle  \Psi^{0,0}_{p,l } \, \mathcal{P}\, \Psi^{0,0}_{p,l}\rangle  \, \, \, \forall \, p, l \,: \, p+l =k+q$. 

From these considerations immediately follows that a dipole-conserving evolution in the LLL $P_0 V P_0$ commutes with the projected dipole operator in real space $P_0 \mathcal P P_0$, as formally proved in the following. Given the  basis for the two-particle states in the LLL $\{\Psi^0_{i}\}_{i=\{k,q\}  }$ (where the two quantum number $k,q$ where grouped for compactness to one index $i$), we want to  show that  $\langle \Psi^0_i | P_0VP_0 P_0 \mathcal P P_0| \Psi^0_j \rangle \overset{?}= \langle  \Psi^0_i |P_0\mathcal P P_0 P_0  V P_0| \Psi^0_j \rangle$  $\forall \, \,i, j $. We insert a resolution of the identity on both sides $\sum_{s}\langle \Psi^0_i | V |\Psi^0_s\rangle \langle \Psi^0_s | \mathcal P | \Psi^0_j \rangle \overset{?} =\sum_s\langle  \Psi^0_i |\mathcal P |\Psi^0_s\rangle \langle \Psi^0_s  | V | \Psi^0_j \rangle$; by using our previous considerations, we find for the left hand side  $\langle\Psi^0_s | \mathcal P | \Psi^0_j \rangle \propto \delta_{s,j}$ and for the right hand side $\langle\Psi^0_i | \mathcal P | \Psi^0_s \rangle \propto \delta_{i,s}$. Then  $\langle \Psi^0_i | V |\Psi^0_j\rangle \langle \Psi^0_j | \mathcal P | \Psi^0_j \rangle \overset{?} =\langle  \Psi^0_i |\mathcal P |\Psi^0_i\rangle \langle \Psi^0_i  | V | \Psi^0_j \rangle$. 
Two cases should be distinguished. On the one hand, when the two states are not connected by the interactions, $\langle\Psi^0_i  | V | \Psi^0_j \rangle = 0$, the relation is trivially fullfilled. On the other hand, when $\langle\Psi^0_i  | V | \Psi^0_j \rangle \neq 0$, then since the interactions $V$ conserves the dipole moment in the LLL, we must have $\langle  \Psi^0_i |\mathcal P |\Psi^0_i\rangle =\langle  \Psi^0_j |\mathcal P |\Psi^0_j\rangle$, which concludes our proof. 

This same analysis can be generalized to the fluctuations of the dipole operator and higher moments; indeed by following the same steps outlined above one easily finds that $P_0\mathcal P^mP_0$ commutes with the projected interactions for every $m$.

As a last point, we notice that the average dipole moment projected in the LLL, $P_0\mathcal P P_0$ coincide with the dipole operator in the LLL.
\bibliography{biblio} 

%apsrev4-2.bst 2019-01-14 (MD) hand-edited version of apsrev4-1.bst
%Control: key (0)
%Control: author (8) initials jnrlst
%Control: editor formatted (1) identically to author
%Control: production of article title (0) allowed
%Control: page (0) single
%Control: year (1) truncated
%Control: production of eprint (0) enabled
\begin{thebibliography}{71}%
\makeatletter
\providecommand \@ifxundefined [1]{%
 \@ifx{#1\undefined}
}%
\providecommand \@ifnum [1]{%
 \ifnum #1\expandafter \@firstoftwo
 \else \expandafter \@secondoftwo
 \fi
}%
\providecommand \@ifx [1]{%
 \ifx #1\expandafter \@firstoftwo
 \else \expandafter \@secondoftwo
 \fi
}%
\providecommand \natexlab [1]{#1}%
\providecommand \enquote  [1]{``#1''}%
\providecommand \bibnamefont  [1]{#1}%
\providecommand \bibfnamefont [1]{#1}%
\providecommand \citenamefont [1]{#1}%
\providecommand \href@noop [0]{\@secondoftwo}%
\providecommand \href [0]{\begingroup \@sanitize@url \@href}%
\providecommand \@href[1]{\@@startlink{#1}\@@href}%
\providecommand \@@href[1]{\endgroup#1\@@endlink}%
\providecommand \@sanitize@url [0]{\catcode `\\12\catcode `\$12\catcode
  `\&12\catcode `\#12\catcode `\^12\catcode `\_12\catcode `\%12\relax}%
\providecommand \@@startlink[1]{}%
\providecommand \@@endlink[0]{}%
\providecommand \url  [0]{\begingroup\@sanitize@url \@url }%
\providecommand \@url [1]{\endgroup\@href {#1}{\urlprefix }}%
\providecommand \urlprefix  [0]{URL }%
\providecommand \Eprint [0]{\href }%
\providecommand \doibase [0]{https://doi.org/}%
\providecommand \selectlanguage [0]{\@gobble}%
\providecommand \bibinfo  [0]{\@secondoftwo}%
\providecommand \bibfield  [0]{\@secondoftwo}%
\providecommand \translation [1]{[#1]}%
\providecommand \BibitemOpen [0]{}%
\providecommand \bibitemStop [0]{}%
\providecommand \bibitemNoStop [0]{.\EOS\space}%
\providecommand \EOS [0]{\spacefactor3000\relax}%
\providecommand \BibitemShut  [1]{\csname bibitem#1\endcsname}%
\let\auto@bib@innerbib\@empty
%</preamble>
\bibitem [{\citenamefont {Tsui}\ \emph {et~al.}(1982)\citenamefont {Tsui},
  \citenamefont {Stormer},\ and\ \citenamefont {Gossard}}]{Tsui1982}%
  \BibitemOpen
  \bibfield  {author} {\bibinfo {author} {\bibfnamefont {D.~C.}\ \bibnamefont
  {Tsui}}, \bibinfo {author} {\bibfnamefont {H.~L.}\ \bibnamefont {Stormer}},\
  and\ \bibinfo {author} {\bibfnamefont {A.~C.}\ \bibnamefont {Gossard}},\
  }\bibfield  {title} {\bibinfo {title} {Two-dimensional magnetotransport in
  the extreme quantum limit},\ }\href
  {https://doi.org/10.1103/PhysRevLett.48.1559} {\bibfield  {journal} {\bibinfo
   {journal} {Phys. Rev. Lett.}\ }\textbf {\bibinfo {volume} {48}},\ \bibinfo
  {pages} {1559} (\bibinfo {year} {1982})}\BibitemShut {NoStop}%
\bibitem [{\citenamefont {Laughlin}(1983)}]{Laughlin83}%
  \BibitemOpen
  \bibfield  {author} {\bibinfo {author} {\bibfnamefont {R.~B.}\ \bibnamefont
  {Laughlin}},\ }\bibfield  {title} {\bibinfo {title} {Anomalous quantum hall
  effect: An incompressible quantum fluid with fractionally charged
  excitations},\ }\href {https://doi.org/10.1103/PhysRevLett.50.1395}
  {\bibfield  {journal} {\bibinfo  {journal} {Phys. Rev. Lett.}\ }\textbf
  {\bibinfo {volume} {50}},\ \bibinfo {pages} {1395} (\bibinfo {year}
  {1983})}\BibitemShut {NoStop}%
\bibitem [{\citenamefont {Matthews}\ \emph {et~al.}(1999)\citenamefont
  {Matthews}, \citenamefont {Anderson}, \citenamefont {Haljan}, \citenamefont
  {Hall}, \citenamefont {Wieman},\ and\ \citenamefont
  {Cornell}}]{Matthews1999}%
  \BibitemOpen
  \bibfield  {author} {\bibinfo {author} {\bibfnamefont {M.~R.}\ \bibnamefont
  {Matthews}}, \bibinfo {author} {\bibfnamefont {B.~P.}\ \bibnamefont
  {Anderson}}, \bibinfo {author} {\bibfnamefont {P.~C.}\ \bibnamefont
  {Haljan}}, \bibinfo {author} {\bibfnamefont {D.~S.}\ \bibnamefont {Hall}},
  \bibinfo {author} {\bibfnamefont {C.~E.}\ \bibnamefont {Wieman}},\ and\
  \bibinfo {author} {\bibfnamefont {E.~A.}\ \bibnamefont {Cornell}},\
  }\bibfield  {title} {\bibinfo {title} {Vortices in a bose-einstein
  condensate},\ }\href {https://doi.org/10.1103/PhysRevLett.83.2498} {\bibfield
   {journal} {\bibinfo  {journal} {Phys. Rev. Lett.}\ }\textbf {\bibinfo
  {volume} {83}},\ \bibinfo {pages} {2498} (\bibinfo {year}
  {1999})}\BibitemShut {NoStop}%
\bibitem [{\citenamefont {Madison}\ \emph {et~al.}(2000)\citenamefont
  {Madison}, \citenamefont {Chevy}, \citenamefont {Wohlleben},\ and\
  \citenamefont {Dalibard}}]{Madison2000}%
  \BibitemOpen
  \bibfield  {author} {\bibinfo {author} {\bibfnamefont {K.~W.}\ \bibnamefont
  {Madison}}, \bibinfo {author} {\bibfnamefont {F.}~\bibnamefont {Chevy}},
  \bibinfo {author} {\bibfnamefont {W.}~\bibnamefont {Wohlleben}},\ and\
  \bibinfo {author} {\bibfnamefont {J.}~\bibnamefont {Dalibard}},\ }\bibfield
  {title} {\bibinfo {title} {Vortex formation in a stirred bose-einstein
  condensate},\ }\href {https://doi.org/10.1103/PhysRevLett.84.806} {\bibfield
  {journal} {\bibinfo  {journal} {Phys. Rev. Lett.}\ }\textbf {\bibinfo
  {volume} {84}},\ \bibinfo {pages} {806} (\bibinfo {year} {2000})}\BibitemShut
  {NoStop}%
\bibitem [{\citenamefont {Abo-Shaeer}\ \emph {et~al.}(2001)\citenamefont
  {Abo-Shaeer}, \citenamefont {Raman}, \citenamefont {Vogels},\ and\
  \citenamefont {Ketterle}}]{Abo-Shaeer2001}%
  \BibitemOpen
  \bibfield  {author} {\bibinfo {author} {\bibfnamefont {J.~R.}\ \bibnamefont
  {Abo-Shaeer}}, \bibinfo {author} {\bibfnamefont {C.}~\bibnamefont {Raman}},
  \bibinfo {author} {\bibfnamefont {J.~M.}\ \bibnamefont {Vogels}},\ and\
  \bibinfo {author} {\bibfnamefont {W.}~\bibnamefont {Ketterle}},\ }\bibfield
  {title} {\bibinfo {title} {{Observation of Vortex Lattices in Bose-Einstein
  Condensates}},\ }\href {https://doi.org/10.1126/science.1060182} {\bibfield
  {journal} {\bibinfo  {journal} {Science}\ }\textbf {\bibinfo {volume}
  {292}},\ \bibinfo {pages} {476} (\bibinfo {year} {2001})}\BibitemShut
  {NoStop}%
\bibitem [{\citenamefont {Schweikhard}\ \emph {et~al.}(2004)\citenamefont
  {Schweikhard}, \citenamefont {Coddington}, \citenamefont {Engels},
  \citenamefont {Mogendorff},\ and\ \citenamefont {Cornell}}]{Schweikhard2004}%
  \BibitemOpen
  \bibfield  {author} {\bibinfo {author} {\bibfnamefont {V.}~\bibnamefont
  {Schweikhard}}, \bibinfo {author} {\bibfnamefont {I.}~\bibnamefont
  {Coddington}}, \bibinfo {author} {\bibfnamefont {P.}~\bibnamefont {Engels}},
  \bibinfo {author} {\bibfnamefont {V.~P.}\ \bibnamefont {Mogendorff}},\ and\
  \bibinfo {author} {\bibfnamefont {E.~A.}\ \bibnamefont {Cornell}},\
  }\bibfield  {title} {\bibinfo {title} {Rapidly rotating bose-einstein
  condensates in and near the lowest landau level},\ }\href
  {https://doi.org/10.1103/PhysRevLett.92.040404} {\bibfield  {journal}
  {\bibinfo  {journal} {Phys. Rev. Lett.}\ }\textbf {\bibinfo {volume} {92}},\
  \bibinfo {pages} {040404} (\bibinfo {year} {2004})}\BibitemShut {NoStop}%
\bibitem [{\citenamefont {Zwierlein}\ \emph {et~al.}(2005)\citenamefont
  {Zwierlein}, \citenamefont {Abo-Shaeer}, \citenamefont {Schirotzek},
  \citenamefont {Schunck},\ and\ \citenamefont {Ketterle}}]{Zwierlein2005}%
  \BibitemOpen
  \bibfield  {author} {\bibinfo {author} {\bibfnamefont {M.~W.}\ \bibnamefont
  {Zwierlein}}, \bibinfo {author} {\bibfnamefont {J.~R.}\ \bibnamefont
  {Abo-Shaeer}}, \bibinfo {author} {\bibfnamefont {A.}~\bibnamefont
  {Schirotzek}}, \bibinfo {author} {\bibfnamefont {C.~H.}\ \bibnamefont
  {Schunck}},\ and\ \bibinfo {author} {\bibfnamefont {W.}~\bibnamefont
  {Ketterle}},\ }\bibfield  {title} {\bibinfo {title} {{Vortices and
  superfluidity in a strongly interacting Fermi gas}},\ }\href
  {https://doi.org/10.1038/nature03858} {\bibfield  {journal} {\bibinfo
  {journal} {Nature}\ }\textbf {\bibinfo {volume} {435}},\ \bibinfo {pages}
  {1047} (\bibinfo {year} {2005})}\BibitemShut {NoStop}%
\bibitem [{\citenamefont {Fletcher}\ \emph {et~al.}(2021)\citenamefont
  {Fletcher}, \citenamefont {Shaffer}, \citenamefont {Wilson}, \citenamefont
  {Patel}, \citenamefont {Yan}, \citenamefont
  {Cr{\ifmmode\acute{e}\else\'{e}\fi}pel}, \citenamefont {Mukherjee},\ and\
  \citenamefont {Zwierlein}}]{Fletcher2021Jun}%
  \BibitemOpen
  \bibfield  {author} {\bibinfo {author} {\bibfnamefont {R.~J.}\ \bibnamefont
  {Fletcher}}, \bibinfo {author} {\bibfnamefont {A.}~\bibnamefont {Shaffer}},
  \bibinfo {author} {\bibfnamefont {C.~C.}\ \bibnamefont {Wilson}}, \bibinfo
  {author} {\bibfnamefont {P.~B.}\ \bibnamefont {Patel}}, \bibinfo {author}
  {\bibfnamefont {Z.}~\bibnamefont {Yan}}, \bibinfo {author} {\bibfnamefont
  {V.}~\bibnamefont {Cr{\ifmmode\acute{e}\else\'{e}\fi}pel}}, \bibinfo {author}
  {\bibfnamefont {B.}~\bibnamefont {Mukherjee}},\ and\ \bibinfo {author}
  {\bibfnamefont {M.~W.}\ \bibnamefont {Zwierlein}},\ }\bibfield  {title}
  {\bibinfo {title} {{Geometric squeezing into the lowest Landau level}},\
  }\href {https://doi.org/10.1126/science.aba7202} {\bibfield  {journal}
  {\bibinfo  {journal} {Science}\ }\textbf {\bibinfo {volume} {372}},\ \bibinfo
  {pages} {1318} (\bibinfo {year} {2021})}\BibitemShut {NoStop}%
\bibitem [{\citenamefont {Mukherjee}\ \emph {et~al.}(2022)\citenamefont
  {Mukherjee}, \citenamefont {Shaffer}, \citenamefont {Patel}, \citenamefont
  {Yan}, \citenamefont {Wilson}, \citenamefont
  {Cr{\ifmmode\acute{e}\else\'{e}\fi}pel}, \citenamefont {Fletcher},\ and\
  \citenamefont {Zwierlein}}]{Mukherjee2022Jan}%
  \BibitemOpen
  \bibfield  {author} {\bibinfo {author} {\bibfnamefont {B.}~\bibnamefont
  {Mukherjee}}, \bibinfo {author} {\bibfnamefont {A.}~\bibnamefont {Shaffer}},
  \bibinfo {author} {\bibfnamefont {P.~B.}\ \bibnamefont {Patel}}, \bibinfo
  {author} {\bibfnamefont {Z.}~\bibnamefont {Yan}}, \bibinfo {author}
  {\bibfnamefont {C.~C.}\ \bibnamefont {Wilson}}, \bibinfo {author}
  {\bibfnamefont {V.}~\bibnamefont {Cr{\ifmmode\acute{e}\else\'{e}\fi}pel}},
  \bibinfo {author} {\bibfnamefont {R.~J.}\ \bibnamefont {Fletcher}},\ and\
  \bibinfo {author} {\bibfnamefont {M.}~\bibnamefont {Zwierlein}},\ }\bibfield
  {title} {\bibinfo {title} {{Crystallization of bosonic quantum Hall states in
  a rotating quantum gas}},\ }\href
  {https://doi.org/10.1038/s41586-021-04170-2} {\bibfield  {journal} {\bibinfo
  {journal} {Nature}\ }\textbf {\bibinfo {volume} {601}},\ \bibinfo {pages}
  {58} (\bibinfo {year} {2022})}\BibitemShut {NoStop}%
\bibitem [{\citenamefont {Yao}\ \emph {et~al.}(2023)\citenamefont {Yao},
  \citenamefont {Chi}, \citenamefont {Mukherjee}, \citenamefont {Shaffer},
  \citenamefont {Zwierlein},\ and\ \citenamefont {Fletcher}}]{Yao2023}%
  \BibitemOpen
  \bibfield  {author} {\bibinfo {author} {\bibfnamefont {R.}~\bibnamefont
  {Yao}}, \bibinfo {author} {\bibfnamefont {S.}~\bibnamefont {Chi}}, \bibinfo
  {author} {\bibfnamefont {B.}~\bibnamefont {Mukherjee}}, \bibinfo {author}
  {\bibfnamefont {A.}~\bibnamefont {Shaffer}}, \bibinfo {author} {\bibfnamefont
  {M.}~\bibnamefont {Zwierlein}},\ and\ \bibinfo {author} {\bibfnamefont
  {R.~J.}\ \bibnamefont {Fletcher}},\ }\href@noop {} {\bibinfo {title}
  {Observation of chiral edge transport in a rapidly-rotating quantum gas}}
  (\bibinfo {year} {2023}),\ \Eprint {https://arxiv.org/abs/2304.10468}
  {arXiv:2304.10468 [cond-mat.quant-gas]} \BibitemShut {NoStop}%
\bibitem [{\citenamefont {Cr{\ifmmode\acute{e}\else\'{e}\fi}pel}\ \emph
  {et~al.}(2023)\citenamefont {Cr{\ifmmode\acute{e}\else\'{e}\fi}pel},
  \citenamefont {Yao}, \citenamefont {Mukherjee}, \citenamefont {Fletcher},\
  and\ \citenamefont {Zwierlein}}]{Crepel2023}%
  \BibitemOpen
  \bibfield  {author} {\bibinfo {author} {\bibfnamefont {V.}~\bibnamefont
  {Cr{\ifmmode\acute{e}\else\'{e}\fi}pel}}, \bibinfo {author} {\bibfnamefont
  {R.}~\bibnamefont {Yao}}, \bibinfo {author} {\bibfnamefont {B.}~\bibnamefont
  {Mukherjee}}, \bibinfo {author} {\bibfnamefont {R.}~\bibnamefont
  {Fletcher}},\ and\ \bibinfo {author} {\bibfnamefont {M.}~\bibnamefont
  {Zwierlein}},\ }\bibfield  {title} {\bibinfo {title} {{Geometric squeezing of
  rotating quantum gases into the lowest Landau level}},\ }\href
  {https://doi.org/10.5802/crphys.173} {\bibfield  {journal} {\bibinfo
  {journal} {C. R. Phys.}\ }\textbf {\bibinfo {volume} {24}},\ \bibinfo {pages}
  {241} (\bibinfo {year} {2023})}\BibitemShut {NoStop}%
\bibitem [{\citenamefont {Lunt}\ \emph {et~al.}(2024)\citenamefont {Lunt},
  \citenamefont {Hill}, \citenamefont {Reiter}, \citenamefont {Preiss},
  \citenamefont {Ga\l{}ka},\ and\ \citenamefont {Jochim}}]{Lunt2024}%
  \BibitemOpen
  \bibfield  {author} {\bibinfo {author} {\bibfnamefont {P.}~\bibnamefont
  {Lunt}}, \bibinfo {author} {\bibfnamefont {P.}~\bibnamefont {Hill}}, \bibinfo
  {author} {\bibfnamefont {J.}~\bibnamefont {Reiter}}, \bibinfo {author}
  {\bibfnamefont {P.~M.}\ \bibnamefont {Preiss}}, \bibinfo {author}
  {\bibfnamefont {M.}~\bibnamefont {Ga\l{}ka}},\ and\ \bibinfo {author}
  {\bibfnamefont {S.}~\bibnamefont {Jochim}},\ }\bibfield  {title} {\bibinfo
  {title} {Realization of a laughlin state of two rapidly rotating fermions},\
  }\href {https://doi.org/10.1103/PhysRevLett.133.253401} {\bibfield  {journal}
  {\bibinfo  {journal} {Phys. Rev. Lett.}\ }\textbf {\bibinfo {volume} {133}},\
  \bibinfo {pages} {253401} (\bibinfo {year} {2024})}\BibitemShut {NoStop}%
\bibitem [{\citenamefont {Lin}\ \emph {et~al.}(2009{\natexlab{a}})\citenamefont
  {Lin}, \citenamefont {Compton}, \citenamefont
  {Jim{\ifmmode\acute{e}\else\'{e}\fi}nez-Garc{\ifmmode\acute{\imath}\else\'{\i}\fi}a},
  \citenamefont {Porto},\ and\ \citenamefont {Spielman}}]{Lin2009Dec}%
  \BibitemOpen
  \bibfield  {author} {\bibinfo {author} {\bibfnamefont {Y.-J.}\ \bibnamefont
  {Lin}}, \bibinfo {author} {\bibfnamefont {R.~L.}\ \bibnamefont {Compton}},
  \bibinfo {author} {\bibfnamefont {K.}~\bibnamefont
  {Jim{\ifmmode\acute{e}\else\'{e}\fi}nez-Garc{\ifmmode\acute{\imath}\else\'{\i}\fi}a}},
  \bibinfo {author} {\bibfnamefont {J.~V.}\ \bibnamefont {Porto}},\ and\
  \bibinfo {author} {\bibfnamefont {I.~B.}\ \bibnamefont {Spielman}},\
  }\bibfield  {title} {\bibinfo {title} {{Synthetic magnetic fields for
  ultracold neutral atoms}},\ }\href {https://doi.org/10.1038/nature08609}
  {\bibfield  {journal} {\bibinfo  {journal} {Nature}\ }\textbf {\bibinfo
  {volume} {462}},\ \bibinfo {pages} {628} (\bibinfo {year}
  {2009}{\natexlab{a}})}\BibitemShut {NoStop}%
\bibitem [{\citenamefont {Lin}\ \emph {et~al.}(2009{\natexlab{b}})\citenamefont
  {Lin}, \citenamefont {Compton}, \citenamefont {Perry}, \citenamefont
  {Phillips}, \citenamefont {Porto},\ and\ \citenamefont {Spielman}}]{Lin2009}%
  \BibitemOpen
  \bibfield  {author} {\bibinfo {author} {\bibfnamefont {Y.-J.}\ \bibnamefont
  {Lin}}, \bibinfo {author} {\bibfnamefont {R.~L.}\ \bibnamefont {Compton}},
  \bibinfo {author} {\bibfnamefont {A.~R.}\ \bibnamefont {Perry}}, \bibinfo
  {author} {\bibfnamefont {W.~D.}\ \bibnamefont {Phillips}}, \bibinfo {author}
  {\bibfnamefont {J.~V.}\ \bibnamefont {Porto}},\ and\ \bibinfo {author}
  {\bibfnamefont {I.~B.}\ \bibnamefont {Spielman}},\ }\bibfield  {title}
  {\bibinfo {title} {Bose-einstein condensate in a uniform light-induced vector
  potential},\ }\href {https://doi.org/10.1103/PhysRevLett.102.130401}
  {\bibfield  {journal} {\bibinfo  {journal} {Phys. Rev. Lett.}\ }\textbf
  {\bibinfo {volume} {102}},\ \bibinfo {pages} {130401} (\bibinfo {year}
  {2009}{\natexlab{b}})}\BibitemShut {NoStop}%
\bibitem [{\citenamefont {Chalopin}\ \emph {et~al.}(2020)\citenamefont
  {Chalopin}, \citenamefont {Satoor}, \citenamefont {Evrard}, \citenamefont
  {Makhalov}, \citenamefont {Dalibard}, \citenamefont {Lopes},\ and\
  \citenamefont {Nascimbene}}]{Chalopin2020Oct}%
  \BibitemOpen
  \bibfield  {author} {\bibinfo {author} {\bibfnamefont {T.}~\bibnamefont
  {Chalopin}}, \bibinfo {author} {\bibfnamefont {T.}~\bibnamefont {Satoor}},
  \bibinfo {author} {\bibfnamefont {A.}~\bibnamefont {Evrard}}, \bibinfo
  {author} {\bibfnamefont {V.}~\bibnamefont {Makhalov}}, \bibinfo {author}
  {\bibfnamefont {J.}~\bibnamefont {Dalibard}}, \bibinfo {author}
  {\bibfnamefont {R.}~\bibnamefont {Lopes}},\ and\ \bibinfo {author}
  {\bibfnamefont {S.}~\bibnamefont {Nascimbene}},\ }\bibfield  {title}
  {\bibinfo {title} {{Probing chiral edge dynamics and bulk topology of a
  synthetic Hall system}},\ }\href {https://doi.org/10.1038/s41567-020-0942-5}
  {\bibfield  {journal} {\bibinfo  {journal} {Nat. Phys.}\ }\textbf {\bibinfo
  {volume} {16}},\ \bibinfo {pages} {1017} (\bibinfo {year}
  {2020})}\BibitemShut {NoStop}%
\bibitem [{\citenamefont {Struck}\ \emph {et~al.}(2012)\citenamefont {Struck},
  \citenamefont {\"Olschl\"ager}, \citenamefont {Weinberg}, \citenamefont
  {Hauke}, \citenamefont {Simonet}, \citenamefont {Eckardt}, \citenamefont
  {Lewenstein}, \citenamefont {Sengstock},\ and\ \citenamefont
  {Windpassinger}}]{Struck2012}%
  \BibitemOpen
  \bibfield  {author} {\bibinfo {author} {\bibfnamefont {J.}~\bibnamefont
  {Struck}}, \bibinfo {author} {\bibfnamefont {C.}~\bibnamefont
  {\"Olschl\"ager}}, \bibinfo {author} {\bibfnamefont {M.}~\bibnamefont
  {Weinberg}}, \bibinfo {author} {\bibfnamefont {P.}~\bibnamefont {Hauke}},
  \bibinfo {author} {\bibfnamefont {J.}~\bibnamefont {Simonet}}, \bibinfo
  {author} {\bibfnamefont {A.}~\bibnamefont {Eckardt}}, \bibinfo {author}
  {\bibfnamefont {M.}~\bibnamefont {Lewenstein}}, \bibinfo {author}
  {\bibfnamefont {K.}~\bibnamefont {Sengstock}},\ and\ \bibinfo {author}
  {\bibfnamefont {P.}~\bibnamefont {Windpassinger}},\ }\bibfield  {title}
  {\bibinfo {title} {Tunable gauge potential for neutral and spinless particles
  in driven optical lattices},\ }\href
  {https://doi.org/10.1103/PhysRevLett.108.225304} {\bibfield  {journal}
  {\bibinfo  {journal} {Phys. Rev. Lett.}\ }\textbf {\bibinfo {volume} {108}},\
  \bibinfo {pages} {225304} (\bibinfo {year} {2012})}\BibitemShut {NoStop}%
\bibitem [{\citenamefont {Aidelsburger}\ \emph {et~al.}(2013)\citenamefont
  {Aidelsburger}, \citenamefont {Atala}, \citenamefont {Lohse}, \citenamefont
  {Barreiro}, \citenamefont {Paredes},\ and\ \citenamefont
  {Bloch}}]{Aidelsburger2013}%
  \BibitemOpen
  \bibfield  {author} {\bibinfo {author} {\bibfnamefont {M.}~\bibnamefont
  {Aidelsburger}}, \bibinfo {author} {\bibfnamefont {M.}~\bibnamefont {Atala}},
  \bibinfo {author} {\bibfnamefont {M.}~\bibnamefont {Lohse}}, \bibinfo
  {author} {\bibfnamefont {J.~T.}\ \bibnamefont {Barreiro}}, \bibinfo {author}
  {\bibfnamefont {B.}~\bibnamefont {Paredes}},\ and\ \bibinfo {author}
  {\bibfnamefont {I.}~\bibnamefont {Bloch}},\ }\bibfield  {title} {\bibinfo
  {title} {Realization of the {Hofstadter} {Hamiltonian} with {Ultracold}
  {Atoms} in {Optical} {Lattices}},\ }\href
  {https://doi.org/10.1103/PhysRevLett.111.185301} {\bibfield  {journal}
  {\bibinfo  {journal} {Phys. Rev. Lett.}\ }\textbf {\bibinfo {volume} {111}},\
  \bibinfo {pages} {185301} (\bibinfo {year} {2013})}\BibitemShut {NoStop}%
\bibitem [{\citenamefont {Miyake}\ \emph {et~al.}(2013)\citenamefont {Miyake},
  \citenamefont {Siviloglou}, \citenamefont {Kennedy}, \citenamefont {Burton},\
  and\ \citenamefont {Ketterle}}]{Miyake2013}%
  \BibitemOpen
  \bibfield  {author} {\bibinfo {author} {\bibfnamefont {H.}~\bibnamefont
  {Miyake}}, \bibinfo {author} {\bibfnamefont {G.~A.}\ \bibnamefont
  {Siviloglou}}, \bibinfo {author} {\bibfnamefont {C.~J.}\ \bibnamefont
  {Kennedy}}, \bibinfo {author} {\bibfnamefont {W.~C.}\ \bibnamefont
  {Burton}},\ and\ \bibinfo {author} {\bibfnamefont {W.}~\bibnamefont
  {Ketterle}},\ }\bibfield  {title} {\bibinfo {title} {Realizing the harper
  hamiltonian with laser-assisted tunneling in optical lattices},\ }\href
  {https://doi.org/10.1103/PhysRevLett.111.185302} {\bibfield  {journal}
  {\bibinfo  {journal} {Phys. Rev. Lett.}\ }\textbf {\bibinfo {volume} {111}},\
  \bibinfo {pages} {185302} (\bibinfo {year} {2013})}\BibitemShut {NoStop}%
\bibitem [{\citenamefont {Jotzu}\ \emph {et~al.}(2014)\citenamefont {Jotzu},
  \citenamefont {Messer}, \citenamefont {Desbuquois}, \citenamefont {Lebrat},
  \citenamefont {Uehlinger}, \citenamefont {Greif},\ and\ \citenamefont
  {Esslinger}}]{Jotzu2014}%
  \BibitemOpen
  \bibfield  {author} {\bibinfo {author} {\bibfnamefont {G.}~\bibnamefont
  {Jotzu}}, \bibinfo {author} {\bibfnamefont {M.}~\bibnamefont {Messer}},
  \bibinfo {author} {\bibfnamefont {R.}~\bibnamefont {Desbuquois}}, \bibinfo
  {author} {\bibfnamefont {M.}~\bibnamefont {Lebrat}}, \bibinfo {author}
  {\bibfnamefont {T.}~\bibnamefont {Uehlinger}}, \bibinfo {author}
  {\bibfnamefont {D.}~\bibnamefont {Greif}},\ and\ \bibinfo {author}
  {\bibfnamefont {T.}~\bibnamefont {Esslinger}},\ }\bibfield  {title} {\bibinfo
  {title} {Experimental realization of the topological {Haldane} model with
  ultracold fermions},\ }\href {https://doi.org/10.1038/nature13915} {\bibfield
   {journal} {\bibinfo  {journal} {Nature}\ }\textbf {\bibinfo {volume}
  {515}},\ \bibinfo {pages} {237} (\bibinfo {year} {2014})}\BibitemShut
  {NoStop}%
\bibitem [{\citenamefont {Atala}\ \emph {et~al.}(2014)\citenamefont {Atala},
  \citenamefont {Aidelsburger}, \citenamefont {Lohse}, \citenamefont
  {Barreiro}, \citenamefont {Paredes},\ and\ \citenamefont
  {Bloch}}]{Atala_2014}%
  \BibitemOpen
  \bibfield  {author} {\bibinfo {author} {\bibfnamefont {M.}~\bibnamefont
  {Atala}}, \bibinfo {author} {\bibfnamefont {M.}~\bibnamefont {Aidelsburger}},
  \bibinfo {author} {\bibfnamefont {M.}~\bibnamefont {Lohse}}, \bibinfo
  {author} {\bibfnamefont {J.~T.}\ \bibnamefont {Barreiro}}, \bibinfo {author}
  {\bibfnamefont {B.}~\bibnamefont {Paredes}},\ and\ \bibinfo {author}
  {\bibfnamefont {I.}~\bibnamefont {Bloch}},\ }\bibfield  {title} {\bibinfo
  {title} {Observation of chiral currents with ultracold atoms in bosonic
  ladders},\ }\href {https://doi.org/10.1038/nphys2998} {\bibfield  {journal}
  {\bibinfo  {journal} {Nature Physics}\ }\textbf {\bibinfo {volume} {10}},\
  \bibinfo {pages} {588–593} (\bibinfo {year} {2014})}\BibitemShut {NoStop}%
\bibitem [{\citenamefont {Aidelsburger}\ \emph {et~al.}(2015)\citenamefont
  {Aidelsburger}, \citenamefont {Lohse}, \citenamefont {Schweizer},
  \citenamefont {Atala}, \citenamefont {Barreiro}, \citenamefont
  {Nascimb{\ifmmode\grave{e}\else\`{e}\fi}ne}, \citenamefont {Cooper},
  \citenamefont {Bloch},\ and\ \citenamefont {Goldman}}]{Aidelsburger2015}%
  \BibitemOpen
  \bibfield  {author} {\bibinfo {author} {\bibfnamefont {M.}~\bibnamefont
  {Aidelsburger}}, \bibinfo {author} {\bibfnamefont {M.}~\bibnamefont {Lohse}},
  \bibinfo {author} {\bibfnamefont {C.}~\bibnamefont {Schweizer}}, \bibinfo
  {author} {\bibfnamefont {M.}~\bibnamefont {Atala}}, \bibinfo {author}
  {\bibfnamefont {J.~T.}\ \bibnamefont {Barreiro}}, \bibinfo {author}
  {\bibfnamefont {S.}~\bibnamefont
  {Nascimb{\ifmmode\grave{e}\else\`{e}\fi}ne}}, \bibinfo {author}
  {\bibfnamefont {N.~R.}\ \bibnamefont {Cooper}}, \bibinfo {author}
  {\bibfnamefont {I.}~\bibnamefont {Bloch}},\ and\ \bibinfo {author}
  {\bibfnamefont {N.}~\bibnamefont {Goldman}},\ }\bibfield  {title} {\bibinfo
  {title} {{Measuring the Chern number of Hofstadter bands with ultracold
  bosonic atoms}},\ }\href {https://doi.org/10.1038/nphys3171} {\bibfield
  {journal} {\bibinfo  {journal} {Nat. Phys.}\ }\textbf {\bibinfo {volume}
  {11}},\ \bibinfo {pages} {162} (\bibinfo {year} {2015})}\BibitemShut
  {NoStop}%
\bibitem [{\citenamefont {Stuhl}\ \emph {et~al.}(2015)\citenamefont {Stuhl},
  \citenamefont {Lu}, \citenamefont {Aycock}, \citenamefont {Genkina},\ and\
  \citenamefont {Spielman}}]{Stuhl2015}%
  \BibitemOpen
  \bibfield  {author} {\bibinfo {author} {\bibfnamefont {B.~K.}\ \bibnamefont
  {Stuhl}}, \bibinfo {author} {\bibfnamefont {H.-I.}\ \bibnamefont {Lu}},
  \bibinfo {author} {\bibfnamefont {L.~M.}\ \bibnamefont {Aycock}}, \bibinfo
  {author} {\bibfnamefont {D.}~\bibnamefont {Genkina}},\ and\ \bibinfo {author}
  {\bibfnamefont {I.~B.}\ \bibnamefont {Spielman}},\ }\bibfield  {title}
  {\bibinfo {title} {{Visualizing edge states with an atomic Bose gas in the
  quantum Hall regime}},\ }\href {https://doi.org/10.1126/science.aaa8515}
  {\bibfield  {journal} {\bibinfo  {journal} {Science}\ }\textbf {\bibinfo
  {volume} {349}},\ \bibinfo {pages} {1514} (\bibinfo {year}
  {2015})}\BibitemShut {NoStop}%
\bibitem [{\citenamefont {Mancini}\ \emph {et~al.}(2015)\citenamefont
  {Mancini}, \citenamefont {Pagano}, \citenamefont {Cappellini}, \citenamefont
  {Livi}, \citenamefont {Rider}, \citenamefont {Catani}, \citenamefont {Sias},
  \citenamefont {Zoller}, \citenamefont {Inguscio}, \citenamefont {Dalmonte},\
  and\ \citenamefont {Fallani}}]{Mancini2015}%
  \BibitemOpen
  \bibfield  {author} {\bibinfo {author} {\bibfnamefont {M.}~\bibnamefont
  {Mancini}}, \bibinfo {author} {\bibfnamefont {G.}~\bibnamefont {Pagano}},
  \bibinfo {author} {\bibfnamefont {G.}~\bibnamefont {Cappellini}}, \bibinfo
  {author} {\bibfnamefont {L.}~\bibnamefont {Livi}}, \bibinfo {author}
  {\bibfnamefont {M.}~\bibnamefont {Rider}}, \bibinfo {author} {\bibfnamefont
  {J.}~\bibnamefont {Catani}}, \bibinfo {author} {\bibfnamefont
  {C.}~\bibnamefont {Sias}}, \bibinfo {author} {\bibfnamefont {P.}~\bibnamefont
  {Zoller}}, \bibinfo {author} {\bibfnamefont {M.}~\bibnamefont {Inguscio}},
  \bibinfo {author} {\bibfnamefont {M.}~\bibnamefont {Dalmonte}},\ and\
  \bibinfo {author} {\bibfnamefont {L.}~\bibnamefont {Fallani}},\ }\bibfield
  {title} {\bibinfo {title} {{Observation of chiral edge states with neutral
  fermions in synthetic Hall ribbons}},\ }\href
  {https://doi.org/10.1126/science.aaa8736} {\bibfield  {journal} {\bibinfo
  {journal} {Science}\ }\textbf {\bibinfo {volume} {349}},\ \bibinfo {pages}
  {1510} (\bibinfo {year} {2015})}\BibitemShut {NoStop}%
\bibitem [{\citenamefont {Flaschner}\ \emph {et~al.}(2016)\citenamefont
  {Flaschner}, \citenamefont {Rem}, \citenamefont {Tarnowski}, \citenamefont
  {Vogel}, \citenamefont {Luhmann}, \citenamefont {Sengstock},\ and\
  \citenamefont {Weitenberg}}]{Flaschner2016}%
  \BibitemOpen
  \bibfield  {author} {\bibinfo {author} {\bibfnamefont {N.}~\bibnamefont
  {Flaschner}}, \bibinfo {author} {\bibfnamefont {B.~S.}\ \bibnamefont {Rem}},
  \bibinfo {author} {\bibfnamefont {M.}~\bibnamefont {Tarnowski}}, \bibinfo
  {author} {\bibfnamefont {D.}~\bibnamefont {Vogel}}, \bibinfo {author}
  {\bibfnamefont {D.-S.}\ \bibnamefont {Luhmann}}, \bibinfo {author}
  {\bibfnamefont {K.}~\bibnamefont {Sengstock}},\ and\ \bibinfo {author}
  {\bibfnamefont {C.}~\bibnamefont {Weitenberg}},\ }\bibfield  {title}
  {\bibinfo {title} {Experimental reconstruction of the {Berry} curvature in a
  {Floquet} {Bloch} band},\ }\href {https://doi.org/10.1126/science.aad4568}
  {\bibfield  {journal} {\bibinfo  {journal} {Science}\ }\textbf {\bibinfo
  {volume} {352}},\ \bibinfo {pages} {1091} (\bibinfo {year}
  {2016})}\BibitemShut {NoStop}%
\bibitem [{\citenamefont {Tai}\ \emph {et~al.}(2017)\citenamefont {Tai},
  \citenamefont {Lukin}, \citenamefont {Rispoli}, \citenamefont {Schittko},
  \citenamefont {Menke}, \citenamefont {Borgnia}, \citenamefont {Preiss},
  \citenamefont {Grusdt}, \citenamefont {Kaufman},\ and\ \citenamefont
  {Greiner}}]{Tai2017}%
  \BibitemOpen
  \bibfield  {author} {\bibinfo {author} {\bibfnamefont {M.~E.}\ \bibnamefont
  {Tai}}, \bibinfo {author} {\bibfnamefont {A.}~\bibnamefont {Lukin}}, \bibinfo
  {author} {\bibfnamefont {M.}~\bibnamefont {Rispoli}}, \bibinfo {author}
  {\bibfnamefont {R.}~\bibnamefont {Schittko}}, \bibinfo {author}
  {\bibfnamefont {T.}~\bibnamefont {Menke}}, \bibinfo {author} {\bibfnamefont
  {D.}~\bibnamefont {Borgnia}}, \bibinfo {author} {\bibfnamefont {P.~M.}\
  \bibnamefont {Preiss}}, \bibinfo {author} {\bibfnamefont {F.}~\bibnamefont
  {Grusdt}}, \bibinfo {author} {\bibfnamefont {A.~M.}\ \bibnamefont
  {Kaufman}},\ and\ \bibinfo {author} {\bibfnamefont {M.}~\bibnamefont
  {Greiner}},\ }\bibfield  {title} {\bibinfo {title} {{Microscopy of the
  interacting Harper{\textendash}Hofstadter model in the two-body limit}},\
  }\href {https://doi.org/10.1038/nature22811} {\bibfield  {journal} {\bibinfo
  {journal} {Nature}\ }\textbf {\bibinfo {volume} {546}},\ \bibinfo {pages}
  {519} (\bibinfo {year} {2017})}\BibitemShut {NoStop}%
\bibitem [{\citenamefont {L{\ifmmode\acute{e}\else\'{e}\fi}onard}\ \emph
  {et~al.}(2023)\citenamefont {L{\ifmmode\acute{e}\else\'{e}\fi}onard},
  \citenamefont {Kim}, \citenamefont {Kwan}, \citenamefont {Segura},
  \citenamefont {Grusdt}, \citenamefont {Repellin}, \citenamefont {Goldman},\
  and\ \citenamefont {Greiner}}]{Leonard2023}%
  \BibitemOpen
  \bibfield  {author} {\bibinfo {author} {\bibfnamefont {J.}~\bibnamefont
  {L{\ifmmode\acute{e}\else\'{e}\fi}onard}}, \bibinfo {author} {\bibfnamefont
  {S.}~\bibnamefont {Kim}}, \bibinfo {author} {\bibfnamefont {J.}~\bibnamefont
  {Kwan}}, \bibinfo {author} {\bibfnamefont {P.}~\bibnamefont {Segura}},
  \bibinfo {author} {\bibfnamefont {F.}~\bibnamefont {Grusdt}}, \bibinfo
  {author} {\bibfnamefont {C.}~\bibnamefont {Repellin}}, \bibinfo {author}
  {\bibfnamefont {N.}~\bibnamefont {Goldman}},\ and\ \bibinfo {author}
  {\bibfnamefont {M.}~\bibnamefont {Greiner}},\ }\bibfield  {title} {\bibinfo
  {title} {{Realization of a fractional quantum Hall state with ultracold
  atoms}},\ }\href {https://doi.org/10.1038/s41586-023-06122-4} {\bibfield
  {journal} {\bibinfo  {journal} {Nature}\ }\textbf {\bibinfo {volume} {619}},\
  \bibinfo {pages} {495} (\bibinfo {year} {2023})}\BibitemShut {NoStop}%
\bibitem [{\citenamefont {Roushan}\ \emph {et~al.}(2017)\citenamefont
  {Roushan}, \citenamefont {Neill}, \citenamefont {Megrant}, \citenamefont
  {Chen}, \citenamefont {Babbush}, \citenamefont {Barends}, \citenamefont
  {Campbell}, \citenamefont {Chen}, \citenamefont {Chiaro}, \citenamefont
  {Dunsworth}, \citenamefont {Fowler}, \citenamefont {Jeffrey}, \citenamefont
  {Kelly}, \citenamefont {Lucero}, \citenamefont {Mutus}, \citenamefont
  {O’Malley}, \citenamefont {Neeley}, \citenamefont {Quintana}, \citenamefont
  {Sank}, \citenamefont {Vainsencher}, \citenamefont {Wenner}, \citenamefont
  {White}, \citenamefont {Kapit}, \citenamefont {Neven},\ and\ \citenamefont
  {Martinis}}]{Roushan2017}%
  \BibitemOpen
  \bibfield  {author} {\bibinfo {author} {\bibfnamefont {P.}~\bibnamefont
  {Roushan}}, \bibinfo {author} {\bibfnamefont {C.}~\bibnamefont {Neill}},
  \bibinfo {author} {\bibfnamefont {A.}~\bibnamefont {Megrant}}, \bibinfo
  {author} {\bibfnamefont {Y.}~\bibnamefont {Chen}}, \bibinfo {author}
  {\bibfnamefont {R.}~\bibnamefont {Babbush}}, \bibinfo {author} {\bibfnamefont
  {R.}~\bibnamefont {Barends}}, \bibinfo {author} {\bibfnamefont
  {B.}~\bibnamefont {Campbell}}, \bibinfo {author} {\bibfnamefont
  {Z.}~\bibnamefont {Chen}}, \bibinfo {author} {\bibfnamefont {B.}~\bibnamefont
  {Chiaro}}, \bibinfo {author} {\bibfnamefont {A.}~\bibnamefont {Dunsworth}},
  \bibinfo {author} {\bibfnamefont {A.}~\bibnamefont {Fowler}}, \bibinfo
  {author} {\bibfnamefont {E.}~\bibnamefont {Jeffrey}}, \bibinfo {author}
  {\bibfnamefont {J.}~\bibnamefont {Kelly}}, \bibinfo {author} {\bibfnamefont
  {E.}~\bibnamefont {Lucero}}, \bibinfo {author} {\bibfnamefont
  {J.}~\bibnamefont {Mutus}}, \bibinfo {author} {\bibfnamefont {P.~J.~J.}\
  \bibnamefont {O’Malley}}, \bibinfo {author} {\bibfnamefont
  {M.}~\bibnamefont {Neeley}}, \bibinfo {author} {\bibfnamefont
  {C.}~\bibnamefont {Quintana}}, \bibinfo {author} {\bibfnamefont
  {D.}~\bibnamefont {Sank}}, \bibinfo {author} {\bibfnamefont {A.}~\bibnamefont
  {Vainsencher}}, \bibinfo {author} {\bibfnamefont {J.}~\bibnamefont {Wenner}},
  \bibinfo {author} {\bibfnamefont {T.}~\bibnamefont {White}}, \bibinfo
  {author} {\bibfnamefont {E.}~\bibnamefont {Kapit}}, \bibinfo {author}
  {\bibfnamefont {H.}~\bibnamefont {Neven}},\ and\ \bibinfo {author}
  {\bibfnamefont {J.}~\bibnamefont {Martinis}},\ }\bibfield  {title} {\bibinfo
  {title} {Chiral ground-state currents of interacting photons in a synthetic
  magnetic field},\ }\href {https://doi.org/10.1038/nphys3930} {\bibfield
  {journal} {\bibinfo  {journal} {Nature Physics}\ }\textbf {\bibinfo {volume}
  {13}},\ \bibinfo {pages} {146} (\bibinfo {year} {2017})}\BibitemShut
  {NoStop}%
\bibitem [{\citenamefont {Clark}\ \emph {et~al.}(2020)\citenamefont {Clark},
  \citenamefont {Schine}, \citenamefont {Baum}, \citenamefont {Jia},\ and\
  \citenamefont {Simon}}]{Clark_2020}%
  \BibitemOpen
  \bibfield  {author} {\bibinfo {author} {\bibfnamefont {L.~W.}\ \bibnamefont
  {Clark}}, \bibinfo {author} {\bibfnamefont {N.}~\bibnamefont {Schine}},
  \bibinfo {author} {\bibfnamefont {C.}~\bibnamefont {Baum}}, \bibinfo {author}
  {\bibfnamefont {N.}~\bibnamefont {Jia}},\ and\ \bibinfo {author}
  {\bibfnamefont {J.}~\bibnamefont {Simon}},\ }\bibfield  {title} {\bibinfo
  {title} {Observation of laughlin states made of light},\ }\href
  {https://doi.org/10.1038/s41586-020-2318-5} {\bibfield  {journal} {\bibinfo
  {journal} {Nature}\ }\textbf {\bibinfo {volume} {582}},\ \bibinfo {pages}
  {41–45} (\bibinfo {year} {2020})}\BibitemShut {NoStop}%
\bibitem [{\citenamefont {Wang}\ \emph {et~al.}(2024)\citenamefont {Wang},
  \citenamefont {Liu}, \citenamefont {Chen}, \citenamefont {Chen},
  \citenamefont {Zhao}, \citenamefont {Ying}, \citenamefont {Shang},
  \citenamefont {Wang}, \citenamefont {Huo}, \citenamefont {Peng},
  \citenamefont {Zhu}, \citenamefont {Lu},\ and\ \citenamefont
  {Pan}}]{Wang2024May}%
  \BibitemOpen
  \bibfield  {author} {\bibinfo {author} {\bibfnamefont {C.}~\bibnamefont
  {Wang}}, \bibinfo {author} {\bibfnamefont {F.-M.}\ \bibnamefont {Liu}},
  \bibinfo {author} {\bibfnamefont {M.-C.}\ \bibnamefont {Chen}}, \bibinfo
  {author} {\bibfnamefont {H.}~\bibnamefont {Chen}}, \bibinfo {author}
  {\bibfnamefont {X.-H.}\ \bibnamefont {Zhao}}, \bibinfo {author}
  {\bibfnamefont {C.}~\bibnamefont {Ying}}, \bibinfo {author} {\bibfnamefont
  {Z.-X.}\ \bibnamefont {Shang}}, \bibinfo {author} {\bibfnamefont {J.-W.}\
  \bibnamefont {Wang}}, \bibinfo {author} {\bibfnamefont {Y.-H.}\ \bibnamefont
  {Huo}}, \bibinfo {author} {\bibfnamefont {C.-Z.}\ \bibnamefont {Peng}},
  \bibinfo {author} {\bibfnamefont {X.}~\bibnamefont {Zhu}}, \bibinfo {author}
  {\bibfnamefont {C.-Y.}\ \bibnamefont {Lu}},\ and\ \bibinfo {author}
  {\bibfnamefont {J.-W.}\ \bibnamefont {Pan}},\ }\bibfield  {title} {\bibinfo
  {title} {{Realization of fractional quantum Hall state with interacting
  photons}},\ }\href {https://doi.org/10.1126/science.ado3912} {\bibfield
  {journal} {\bibinfo  {journal} {Science}\ }\textbf {\bibinfo {volume}
  {384}},\ \bibinfo {pages} {579} (\bibinfo {year} {2024})}\BibitemShut
  {NoStop}%
\bibitem [{\citenamefont {Liu}\ \emph {et~al.}(2018)\citenamefont {Liu},
  \citenamefont {Gromov},\ and\ \citenamefont {Papi\ifmmode~\acute{c}\else
  \'{c}\fi{}}}]{Liu2018}%
  \BibitemOpen
  \bibfield  {author} {\bibinfo {author} {\bibfnamefont {Z.}~\bibnamefont
  {Liu}}, \bibinfo {author} {\bibfnamefont {A.}~\bibnamefont {Gromov}},\ and\
  \bibinfo {author} {\bibfnamefont {Z.}~\bibnamefont
  {Papi\ifmmode~\acute{c}\else \'{c}\fi{}}},\ }\bibfield  {title} {\bibinfo
  {title} {Geometric quench and nonequilibrium dynamics of fractional quantum
  hall states},\ }\href {https://doi.org/10.1103/PhysRevB.98.155140} {\bibfield
   {journal} {\bibinfo  {journal} {Phys. Rev. B}\ }\textbf {\bibinfo {volume}
  {98}},\ \bibinfo {pages} {155140} (\bibinfo {year} {2018})}\BibitemShut
  {NoStop}%
\bibitem [{\citenamefont {Fremling}\ \emph {et~al.}(2018)\citenamefont
  {Fremling}, \citenamefont {Repellin}, \citenamefont
  {St{\ifmmode\acute{e}\else\'{e}\fi}phan}, \citenamefont {Moran},
  \citenamefont {Slingerland},\ and\ \citenamefont {Haque}}]{Fremling2018Oct}%
  \BibitemOpen
  \bibfield  {author} {\bibinfo {author} {\bibfnamefont {M.}~\bibnamefont
  {Fremling}}, \bibinfo {author} {\bibfnamefont {C.}~\bibnamefont {Repellin}},
  \bibinfo {author} {\bibfnamefont {J.-M.}\ \bibnamefont
  {St{\ifmmode\acute{e}\else\'{e}\fi}phan}}, \bibinfo {author} {\bibfnamefont
  {N.}~\bibnamefont {Moran}}, \bibinfo {author} {\bibfnamefont {J.~K.}\
  \bibnamefont {Slingerland}},\ and\ \bibinfo {author} {\bibfnamefont
  {M.}~\bibnamefont {Haque}},\ }\bibfield  {title} {\bibinfo {title} {{Dynamics
  and level statistics of interacting fermions in the lowest Landau level}},\
  }\href {https://doi.org/10.1088/1367-2630/aae73f} {\bibfield  {journal}
  {\bibinfo  {journal} {New J. Phys.}\ }\textbf {\bibinfo {volume} {20}},\
  \bibinfo {pages} {103036} (\bibinfo {year} {2018})}\BibitemShut {NoStop}%
\bibitem [{\citenamefont {Liu}\ \emph {et~al.}(2021)\citenamefont {Liu},
  \citenamefont {Balram}, \citenamefont {Papi\ifmmode~\acute{c}\else
  \'{c}\fi{}},\ and\ \citenamefont {Gromov}}]{Liu2021}%
  \BibitemOpen
  \bibfield  {author} {\bibinfo {author} {\bibfnamefont {Z.}~\bibnamefont
  {Liu}}, \bibinfo {author} {\bibfnamefont {A.~C.}\ \bibnamefont {Balram}},
  \bibinfo {author} {\bibfnamefont {Z.}~\bibnamefont
  {Papi\ifmmode~\acute{c}\else \'{c}\fi{}}},\ and\ \bibinfo {author}
  {\bibfnamefont {A.}~\bibnamefont {Gromov}},\ }\bibfield  {title} {\bibinfo
  {title} {Quench dynamics of collective modes in fractional quantum hall
  bilayers},\ }\href {https://doi.org/10.1103/PhysRevLett.126.076604}
  {\bibfield  {journal} {\bibinfo  {journal} {Phys. Rev. Lett.}\ }\textbf
  {\bibinfo {volume} {126}},\ \bibinfo {pages} {076604} (\bibinfo {year}
  {2021})}\BibitemShut {NoStop}%
\bibitem [{\citenamefont {Haldane}(1983)}]{Haldane1983}%
  \BibitemOpen
  \bibfield  {author} {\bibinfo {author} {\bibfnamefont {F.~D.~M.}\
  \bibnamefont {Haldane}},\ }\bibfield  {title} {\bibinfo {title} {Fractional
  quantization of the hall effect: A hierarchy of incompressible quantum fluid
  states},\ }\href {https://doi.org/10.1103/PhysRevLett.51.605} {\bibfield
  {journal} {\bibinfo  {journal} {Phys. Rev. Lett.}\ }\textbf {\bibinfo
  {volume} {51}},\ \bibinfo {pages} {605} (\bibinfo {year} {1983})}\BibitemShut
  {NoStop}%
\bibitem [{\citenamefont {Haldane}(1985)}]{Haldane1985}%
  \BibitemOpen
  \bibfield  {author} {\bibinfo {author} {\bibfnamefont {F.~D.~M.}\
  \bibnamefont {Haldane}},\ }\bibfield  {title} {\bibinfo {title}
  {Many-particle translational symmetries of two-dimensional electrons at
  rational landau-level filling},\ }\href
  {https://doi.org/10.1103/PhysRevLett.55.2095} {\bibfield  {journal} {\bibinfo
   {journal} {Phys. Rev. Lett.}\ }\textbf {\bibinfo {volume} {55}},\ \bibinfo
  {pages} {2095} (\bibinfo {year} {1985})}\BibitemShut {NoStop}%
\bibitem [{\citenamefont {Trugman}\ and\ \citenamefont
  {Kivelson}(1985)}]{Trugman1985}%
  \BibitemOpen
  \bibfield  {author} {\bibinfo {author} {\bibfnamefont {S.~A.}\ \bibnamefont
  {Trugman}}\ and\ \bibinfo {author} {\bibfnamefont {S.}~\bibnamefont
  {Kivelson}},\ }\bibfield  {title} {\bibinfo {title} {Exact results for the
  fractional quantum hall effect with general interactions},\ }\href
  {https://doi.org/10.1103/PhysRevB.31.5280} {\bibfield  {journal} {\bibinfo
  {journal} {Phys. Rev. B}\ }\textbf {\bibinfo {volume} {31}},\ \bibinfo
  {pages} {5280} (\bibinfo {year} {1985})}\BibitemShut {NoStop}%
\bibitem [{\citenamefont {Bergholtz}\ and\ \citenamefont
  {Karlhede}(2005)}]{Bergholtz2005}%
  \BibitemOpen
  \bibfield  {author} {\bibinfo {author} {\bibfnamefont {E.~J.}\ \bibnamefont
  {Bergholtz}}\ and\ \bibinfo {author} {\bibfnamefont {A.}~\bibnamefont
  {Karlhede}},\ }\bibfield  {title} {\bibinfo {title} {Half-filled lowest
  landau level on a thin torus},\ }\href
  {https://doi.org/10.1103/PhysRevLett.94.026802} {\bibfield  {journal}
  {\bibinfo  {journal} {Phys. Rev. Lett.}\ }\textbf {\bibinfo {volume} {94}},\
  \bibinfo {pages} {026802} (\bibinfo {year} {2005})}\BibitemShut {NoStop}%
\bibitem [{\citenamefont {Seidel}\ \emph {et~al.}(2005)\citenamefont {Seidel},
  \citenamefont {Fu}, \citenamefont {Lee}, \citenamefont {Leinaas},\ and\
  \citenamefont {Moore}}]{Alex2005}%
  \BibitemOpen
  \bibfield  {author} {\bibinfo {author} {\bibfnamefont {A.}~\bibnamefont
  {Seidel}}, \bibinfo {author} {\bibfnamefont {H.}~\bibnamefont {Fu}}, \bibinfo
  {author} {\bibfnamefont {D.-H.}\ \bibnamefont {Lee}}, \bibinfo {author}
  {\bibfnamefont {J.~M.}\ \bibnamefont {Leinaas}},\ and\ \bibinfo {author}
  {\bibfnamefont {J.}~\bibnamefont {Moore}},\ }\bibfield  {title} {\bibinfo
  {title} {Incompressible quantum liquids and new conservation laws},\ }\href
  {https://doi.org/10.1103/PhysRevLett.95.266405} {\bibfield  {journal}
  {\bibinfo  {journal} {Phys. Rev. Lett.}\ }\textbf {\bibinfo {volume} {95}},\
  \bibinfo {pages} {266405} (\bibinfo {year} {2005})}\BibitemShut {NoStop}%
\bibitem [{\citenamefont {Nakamura}\ \emph {et~al.}(2012)\citenamefont
  {Nakamura}, \citenamefont {Wang},\ and\ \citenamefont
  {Bergholtz}}]{Nakamura2012}%
  \BibitemOpen
  \bibfield  {author} {\bibinfo {author} {\bibfnamefont {M.}~\bibnamefont
  {Nakamura}}, \bibinfo {author} {\bibfnamefont {Z.-Y.}\ \bibnamefont {Wang}},\
  and\ \bibinfo {author} {\bibfnamefont {E.~J.}\ \bibnamefont {Bergholtz}},\
  }\bibfield  {title} {\bibinfo {title} {Exactly solvable fermion chain
  describing a $\ensuremath{\nu}=1/3$ fractional quantum hall state},\ }\href
  {https://doi.org/10.1103/PhysRevLett.109.016401} {\bibfield  {journal}
  {\bibinfo  {journal} {Phys. Rev. Lett.}\ }\textbf {\bibinfo {volume} {109}},\
  \bibinfo {pages} {016401} (\bibinfo {year} {2012})}\BibitemShut {NoStop}%
\bibitem [{\citenamefont {Bergholtz}\ and\ \citenamefont
  {Karlhede}(2006)}]{Bergholtz2006}%
  \BibitemOpen
  \bibfield  {author} {\bibinfo {author} {\bibfnamefont {E.~J.}\ \bibnamefont
  {Bergholtz}}\ and\ \bibinfo {author} {\bibfnamefont {A.}~\bibnamefont
  {Karlhede}},\ }\bibfield  {title} {\bibinfo {title} {{{`}One-dimensional{'}
  theory of the quantum Hall system}},\ }\href
  {https://doi.org/10.1088/1742-5468/2006/04/L04001} {\bibfield  {journal}
  {\bibinfo  {journal} {J. Stat. Mech.: Theory Exp.}\ }\textbf {\bibinfo
  {volume} {2006}}\bibinfo  {number} { (04)},\ \bibinfo {pages}
  {L04001}}\BibitemShut {NoStop}%
\bibitem [{\citenamefont {Rezayi}\ and\ \citenamefont
  {Haldane}(1994)}]{Rezayi1994}%
  \BibitemOpen
\bibfield  {number} {  }\bibfield  {author} {\bibinfo {author} {\bibfnamefont
  {E.~H.}\ \bibnamefont {Rezayi}}\ and\ \bibinfo {author} {\bibfnamefont
  {F.~D.~M.}\ \bibnamefont {Haldane}},\ }\bibfield  {title} {\bibinfo {title}
  {Laughlin state on stretched and squeezed cylinders and edge excitations in
  the quantum hall effect},\ }\href {https://doi.org/10.1103/PhysRevB.50.17199}
  {\bibfield  {journal} {\bibinfo  {journal} {Phys. Rev. B}\ }\textbf {\bibinfo
  {volume} {50}},\ \bibinfo {pages} {17199} (\bibinfo {year}
  {1994})}\BibitemShut {NoStop}%
\bibitem [{\citenamefont {Bergholtz}\ and\ \citenamefont
  {Karlhede}(2008)}]{Bergholtz2008}%
  \BibitemOpen
  \bibfield  {author} {\bibinfo {author} {\bibfnamefont {E.~J.}\ \bibnamefont
  {Bergholtz}}\ and\ \bibinfo {author} {\bibfnamefont {A.}~\bibnamefont
  {Karlhede}},\ }\bibfield  {title} {\bibinfo {title} {Quantum hall system in
  tao-thouless limit},\ }\href {https://doi.org/10.1103/PhysRevB.77.155308}
  {\bibfield  {journal} {\bibinfo  {journal} {Phys. Rev. B}\ }\textbf {\bibinfo
  {volume} {77}},\ \bibinfo {pages} {155308} (\bibinfo {year}
  {2008})}\BibitemShut {NoStop}%
\bibitem [{\citenamefont {Moudgalya}\ \emph {et~al.}(2020)\citenamefont
  {Moudgalya}, \citenamefont {Bernevig},\ and\ \citenamefont
  {Regnault}}]{Moudgalya2020}%
  \BibitemOpen
  \bibfield  {author} {\bibinfo {author} {\bibfnamefont {S.}~\bibnamefont
  {Moudgalya}}, \bibinfo {author} {\bibfnamefont {B.~A.}\ \bibnamefont
  {Bernevig}},\ and\ \bibinfo {author} {\bibfnamefont {N.}~\bibnamefont
  {Regnault}},\ }\bibfield  {title} {\bibinfo {title} {Quantum many-body scars
  in a landau level on a thin torus},\ }\href
  {https://doi.org/10.1103/PhysRevB.102.195150} {\bibfield  {journal} {\bibinfo
   {journal} {Phys. Rev. B}\ }\textbf {\bibinfo {volume} {102}},\ \bibinfo
  {pages} {195150} (\bibinfo {year} {2020})}\BibitemShut {NoStop}%
\bibitem [{\citenamefont {Nachtergaele}\ \emph {et~al.}(2020)\citenamefont
  {Nachtergaele}, \citenamefont {Warzel},\ and\ \citenamefont
  {Young}}]{Nachtergaele2020Nov}%
  \BibitemOpen
  \bibfield  {author} {\bibinfo {author} {\bibfnamefont {B.}~\bibnamefont
  {Nachtergaele}}, \bibinfo {author} {\bibfnamefont {S.}~\bibnamefont
  {Warzel}},\ and\ \bibinfo {author} {\bibfnamefont {A.}~\bibnamefont
  {Young}},\ }\bibfield  {title} {\bibinfo {title} {{Low-complexity eigenstates
  of a {$\nu$} = 1/3 fractional quantum Hall system}},\ }\href
  {https://doi.org/10.1088/1751-8121/abca73} {\bibfield  {journal} {\bibinfo
  {journal} {J. Phys. A: Math. Theor.}\ }\textbf {\bibinfo {volume} {54}},\
  \bibinfo {pages} {01LT01} (\bibinfo {year} {2020})}\BibitemShut {NoStop}%
\bibitem [{\citenamefont {Gromov}\ \emph {et~al.}(2020)\citenamefont {Gromov},
  \citenamefont {Lucas},\ and\ \citenamefont {Nandkishore}}]{Gromov2020}%
  \BibitemOpen
  \bibfield  {author} {\bibinfo {author} {\bibfnamefont {A.}~\bibnamefont
  {Gromov}}, \bibinfo {author} {\bibfnamefont {A.}~\bibnamefont {Lucas}},\ and\
  \bibinfo {author} {\bibfnamefont {R.~M.}\ \bibnamefont {Nandkishore}},\
  }\bibfield  {title} {\bibinfo {title} {Fracton hydrodynamics},\ }\href
  {https://doi.org/10.1103/PhysRevResearch.2.033124} {\bibfield  {journal}
  {\bibinfo  {journal} {Phys. Rev. Res.}\ }\textbf {\bibinfo {volume} {2}},\
  \bibinfo {pages} {033124} (\bibinfo {year} {2020})}\BibitemShut {NoStop}%
\bibitem [{\citenamefont {Feldmeier}\ \emph {et~al.}(2020)\citenamefont
  {Feldmeier}, \citenamefont {Sala}, \citenamefont {De~Tomasi}, \citenamefont
  {Pollmann},\ and\ \citenamefont {Knap}}]{Feldermeier2020}%
  \BibitemOpen
  \bibfield  {author} {\bibinfo {author} {\bibfnamefont {J.}~\bibnamefont
  {Feldmeier}}, \bibinfo {author} {\bibfnamefont {P.}~\bibnamefont {Sala}},
  \bibinfo {author} {\bibfnamefont {G.}~\bibnamefont {De~Tomasi}}, \bibinfo
  {author} {\bibfnamefont {F.}~\bibnamefont {Pollmann}},\ and\ \bibinfo
  {author} {\bibfnamefont {M.}~\bibnamefont {Knap}},\ }\bibfield  {title}
  {\bibinfo {title} {Anomalous diffusion in dipole- and
  higher-moment-conserving systems},\ }\href
  {https://doi.org/10.1103/PhysRevLett.125.245303} {\bibfield  {journal}
  {\bibinfo  {journal} {Phys. Rev. Lett.}\ }\textbf {\bibinfo {volume} {125}},\
  \bibinfo {pages} {245303} (\bibinfo {year} {2020})}\BibitemShut {NoStop}%
\bibitem [{\citenamefont {Guardado-Sanchez}\ \emph {et~al.}(2020)\citenamefont
  {Guardado-Sanchez}, \citenamefont {Morningstar}, \citenamefont {Spar},
  \citenamefont {Brown}, \citenamefont {Huse},\ and\ \citenamefont
  {Bakr}}]{Guardado2020}%
  \BibitemOpen
  \bibfield  {author} {\bibinfo {author} {\bibfnamefont {E.}~\bibnamefont
  {Guardado-Sanchez}}, \bibinfo {author} {\bibfnamefont {A.}~\bibnamefont
  {Morningstar}}, \bibinfo {author} {\bibfnamefont {B.~M.}\ \bibnamefont
  {Spar}}, \bibinfo {author} {\bibfnamefont {P.~T.}\ \bibnamefont {Brown}},
  \bibinfo {author} {\bibfnamefont {D.~A.}\ \bibnamefont {Huse}},\ and\
  \bibinfo {author} {\bibfnamefont {W.~S.}\ \bibnamefont {Bakr}},\ }\bibfield
  {title} {\bibinfo {title} {Subdiffusion and heat transport in a tilted
  two-dimensional fermi-hubbard system},\ }\href
  {https://doi.org/10.1103/PhysRevX.10.011042} {\bibfield  {journal} {\bibinfo
  {journal} {Phys. Rev. X}\ }\textbf {\bibinfo {volume} {10}},\ \bibinfo
  {pages} {011042} (\bibinfo {year} {2020})}\BibitemShut {NoStop}%
\bibitem [{\citenamefont {Ortiz}\ \emph {et~al.}(2013)\citenamefont {Ortiz},
  \citenamefont {Nussinov}, \citenamefont {Dukelsky},\ and\ \citenamefont
  {Seidel}}]{Ortiz13}%
  \BibitemOpen
  \bibfield  {author} {\bibinfo {author} {\bibfnamefont {G.}~\bibnamefont
  {Ortiz}}, \bibinfo {author} {\bibfnamefont {Z.}~\bibnamefont {Nussinov}},
  \bibinfo {author} {\bibfnamefont {J.}~\bibnamefont {Dukelsky}},\ and\
  \bibinfo {author} {\bibfnamefont {A.}~\bibnamefont {Seidel}},\ }\bibfield
  {title} {\bibinfo {title} {Repulsive interactions in quantum hall systems as
  a pairing problem},\ }\href {https://doi.org/10.1103/PhysRevB.88.165303}
  {\bibfield  {journal} {\bibinfo  {journal} {Phys. Rev. B}\ }\textbf {\bibinfo
  {volume} {88}},\ \bibinfo {pages} {165303} (\bibinfo {year}
  {2013})}\BibitemShut {NoStop}%
\bibitem [{\citenamefont {Scaffidi}\ and\ \citenamefont
  {Simon}(2014)}]{Scaffidi2014}%
  \BibitemOpen
  \bibfield  {author} {\bibinfo {author} {\bibfnamefont {T.}~\bibnamefont
  {Scaffidi}}\ and\ \bibinfo {author} {\bibfnamefont {S.~H.}\ \bibnamefont
  {Simon}},\ }\bibfield  {title} {\bibinfo {title} {Exact solutions of
  fractional chern insulators: Interacting particles in the hofstadter model at
  finite size},\ }\href {https://doi.org/10.1103/PhysRevB.90.115132} {\bibfield
   {journal} {\bibinfo  {journal} {Phys. Rev. B}\ }\textbf {\bibinfo {volume}
  {90}},\ \bibinfo {pages} {115132} (\bibinfo {year} {2014})}\BibitemShut
  {NoStop}%
\bibitem [{\citenamefont {Chen}\ \emph {et~al.}(2014)\citenamefont {Chen},
  \citenamefont {Mazaheri}, \citenamefont {Seidel},\ and\ \citenamefont
  {Tang}}]{Chen2014}%
  \BibitemOpen
  \bibfield  {author} {\bibinfo {author} {\bibfnamefont {L.}~\bibnamefont
  {Chen}}, \bibinfo {author} {\bibfnamefont {T.}~\bibnamefont {Mazaheri}},
  \bibinfo {author} {\bibfnamefont {A.}~\bibnamefont {Seidel}},\ and\ \bibinfo
  {author} {\bibfnamefont {X.}~\bibnamefont {Tang}},\ }\bibfield  {title}
  {\bibinfo {title} {{The impossibility of exactly flat non-trivial Chern bands
  in strictly local periodic tight binding models}},\ }\href
  {https://doi.org/10.1088/1751-8113/47/15/152001} {\bibfield  {journal}
  {\bibinfo  {journal} {J. Phys. A: Math. Theor.}\ }\textbf {\bibinfo {volume}
  {47}},\ \bibinfo {pages} {152001} (\bibinfo {year} {2014})}\BibitemShut
  {NoStop}%
\bibitem [{\citenamefont {S\o{}rensen}\ \emph {et~al.}(2005)\citenamefont
  {S\o{}rensen}, \citenamefont {Demler},\ and\ \citenamefont
  {Lukin}}]{Sorensen2005}%
  \BibitemOpen
  \bibfield  {author} {\bibinfo {author} {\bibfnamefont {A.~S.}\ \bibnamefont
  {S\o{}rensen}}, \bibinfo {author} {\bibfnamefont {E.}~\bibnamefont
  {Demler}},\ and\ \bibinfo {author} {\bibfnamefont {M.~D.}\ \bibnamefont
  {Lukin}},\ }\bibfield  {title} {\bibinfo {title} {Fractional quantum hall
  states of atoms in optical lattices},\ }\href
  {https://doi.org/10.1103/PhysRevLett.94.086803} {\bibfield  {journal}
  {\bibinfo  {journal} {Phys. Rev. Lett.}\ }\textbf {\bibinfo {volume} {94}},\
  \bibinfo {pages} {086803} (\bibinfo {year} {2005})}\BibitemShut {NoStop}%
\bibitem [{\citenamefont {Hafezi}\ \emph {et~al.}(2007)\citenamefont {Hafezi},
  \citenamefont {S\o{}rensen}, \citenamefont {Demler},\ and\ \citenamefont
  {Lukin}}]{Hafezi2007}%
  \BibitemOpen
  \bibfield  {author} {\bibinfo {author} {\bibfnamefont {M.}~\bibnamefont
  {Hafezi}}, \bibinfo {author} {\bibfnamefont {A.~S.}\ \bibnamefont
  {S\o{}rensen}}, \bibinfo {author} {\bibfnamefont {E.}~\bibnamefont
  {Demler}},\ and\ \bibinfo {author} {\bibfnamefont {M.~D.}\ \bibnamefont
  {Lukin}},\ }\bibfield  {title} {\bibinfo {title} {Fractional quantum hall
  effect in optical lattices},\ }\href
  {https://doi.org/10.1103/PhysRevA.76.023613} {\bibfield  {journal} {\bibinfo
  {journal} {Phys. Rev. A}\ }\textbf {\bibinfo {volume} {76}},\ \bibinfo
  {pages} {023613} (\bibinfo {year} {2007})}\BibitemShut {NoStop}%
\bibitem [{\citenamefont {Harper}\ \emph {et~al.}(2014)\citenamefont {Harper},
  \citenamefont {Simon},\ and\ \citenamefont {Roy}}]{Harper2014}%
  \BibitemOpen
  \bibfield  {author} {\bibinfo {author} {\bibfnamefont {F.}~\bibnamefont
  {Harper}}, \bibinfo {author} {\bibfnamefont {S.~H.}\ \bibnamefont {Simon}},\
  and\ \bibinfo {author} {\bibfnamefont {R.}~\bibnamefont {Roy}},\ }\bibfield
  {title} {\bibinfo {title} {Perturbative approach to flat chern bands in the
  hofstadter model},\ }\href {https://doi.org/10.1103/PhysRevB.90.075104}
  {\bibfield  {journal} {\bibinfo  {journal} {Phys. Rev. B}\ }\textbf {\bibinfo
  {volume} {90}},\ \bibinfo {pages} {075104} (\bibinfo {year}
  {2014})}\BibitemShut {NoStop}%
\bibitem [{\citenamefont {Sala}\ \emph {et~al.}(2020)\citenamefont {Sala},
  \citenamefont {Rakovszky}, \citenamefont {Verresen}, \citenamefont {Knap},\
  and\ \citenamefont {Pollmann}}]{Sala2020}%
  \BibitemOpen
  \bibfield  {author} {\bibinfo {author} {\bibfnamefont {P.}~\bibnamefont
  {Sala}}, \bibinfo {author} {\bibfnamefont {T.}~\bibnamefont {Rakovszky}},
  \bibinfo {author} {\bibfnamefont {R.}~\bibnamefont {Verresen}}, \bibinfo
  {author} {\bibfnamefont {M.}~\bibnamefont {Knap}},\ and\ \bibinfo {author}
  {\bibfnamefont {F.}~\bibnamefont {Pollmann}},\ }\bibfield  {title} {\bibinfo
  {title} {Ergodicity breaking arising from hilbert space fragmentation in
  dipole-conserving hamiltonians},\ }\href
  {https://doi.org/10.1103/PhysRevX.10.011047} {\bibfield  {journal} {\bibinfo
  {journal} {Phys. Rev. X}\ }\textbf {\bibinfo {volume} {10}},\ \bibinfo
  {pages} {011047} (\bibinfo {year} {2020})}\BibitemShut {NoStop}%
\bibitem [{\citenamefont {Rakovszky}\ \emph {et~al.}(2020)\citenamefont
  {Rakovszky}, \citenamefont {Sala}, \citenamefont {Verresen}, \citenamefont
  {Knap},\ and\ \citenamefont {Pollmann}}]{Rakovszky2020}%
  \BibitemOpen
  \bibfield  {author} {\bibinfo {author} {\bibfnamefont {T.}~\bibnamefont
  {Rakovszky}}, \bibinfo {author} {\bibfnamefont {P.}~\bibnamefont {Sala}},
  \bibinfo {author} {\bibfnamefont {R.}~\bibnamefont {Verresen}}, \bibinfo
  {author} {\bibfnamefont {M.}~\bibnamefont {Knap}},\ and\ \bibinfo {author}
  {\bibfnamefont {F.}~\bibnamefont {Pollmann}},\ }\bibfield  {title} {\bibinfo
  {title} {Statistical localization: From strong fragmentation to strong edge
  modes},\ }\href {https://doi.org/10.1103/PhysRevB.101.125126} {\bibfield
  {journal} {\bibinfo  {journal} {Phys. Rev. B}\ }\textbf {\bibinfo {volume}
  {101}},\ \bibinfo {pages} {125126} (\bibinfo {year} {2020})}\BibitemShut
  {NoStop}%
\bibitem [{\citenamefont {Khemani}\ \emph {et~al.}(2020)\citenamefont
  {Khemani}, \citenamefont {Hermele},\ and\ \citenamefont
  {Nandkishore}}]{Khemani2020}%
  \BibitemOpen
  \bibfield  {author} {\bibinfo {author} {\bibfnamefont {V.}~\bibnamefont
  {Khemani}}, \bibinfo {author} {\bibfnamefont {M.}~\bibnamefont {Hermele}},\
  and\ \bibinfo {author} {\bibfnamefont {R.}~\bibnamefont {Nandkishore}},\
  }\bibfield  {title} {\bibinfo {title} {Localization from hilbert space
  shattering: From theory to physical realizations},\ }\href
  {https://doi.org/10.1103/PhysRevB.101.174204} {\bibfield  {journal} {\bibinfo
   {journal} {Phys. Rev. B}\ }\textbf {\bibinfo {volume} {101}},\ \bibinfo
  {pages} {174204} (\bibinfo {year} {2020})}\BibitemShut {NoStop}%
\bibitem [{\citenamefont {Morningstar}\ \emph {et~al.}(2020)\citenamefont
  {Morningstar}, \citenamefont {Khemani},\ and\ \citenamefont
  {Huse}}]{Morningstar2020}%
  \BibitemOpen
  \bibfield  {author} {\bibinfo {author} {\bibfnamefont {A.}~\bibnamefont
  {Morningstar}}, \bibinfo {author} {\bibfnamefont {V.}~\bibnamefont
  {Khemani}},\ and\ \bibinfo {author} {\bibfnamefont {D.~A.}\ \bibnamefont
  {Huse}},\ }\bibfield  {title} {\bibinfo {title} {Kinetically constrained
  freezing transition in a dipole-conserving system},\ }\href
  {https://doi.org/10.1103/PhysRevB.101.214205} {\bibfield  {journal} {\bibinfo
   {journal} {Phys. Rev. B}\ }\textbf {\bibinfo {volume} {101}},\ \bibinfo
  {pages} {214205} (\bibinfo {year} {2020})}\BibitemShut {NoStop}%
\bibitem [{\citenamefont {Zechmann}\ \emph {et~al.}(2023)\citenamefont
  {Zechmann}, \citenamefont {Altman}, \citenamefont {Knap},\ and\ \citenamefont
  {Feldmeier}}]{Zechmann23}%
  \BibitemOpen
  \bibfield  {author} {\bibinfo {author} {\bibfnamefont {P.}~\bibnamefont
  {Zechmann}}, \bibinfo {author} {\bibfnamefont {E.}~\bibnamefont {Altman}},
  \bibinfo {author} {\bibfnamefont {M.}~\bibnamefont {Knap}},\ and\ \bibinfo
  {author} {\bibfnamefont {J.}~\bibnamefont {Feldmeier}},\ }\bibfield  {title}
  {\bibinfo {title} {Fractonic luttinger liquids and supersolids in a
  constrained bose-hubbard model},\ }\href
  {https://doi.org/10.1103/PhysRevB.107.195131} {\bibfield  {journal} {\bibinfo
   {journal} {Phys. Rev. B}\ }\textbf {\bibinfo {volume} {107}},\ \bibinfo
  {pages} {195131} (\bibinfo {year} {2023})}\BibitemShut {NoStop}%
\bibitem [{\citenamefont {Lake}\ \emph {et~al.}(2023)\citenamefont {Lake},
  \citenamefont {Lee}, \citenamefont {Han},\ and\ \citenamefont
  {Senthil}}]{Lake_2023}%
  \BibitemOpen
  \bibfield  {author} {\bibinfo {author} {\bibfnamefont {E.}~\bibnamefont
  {Lake}}, \bibinfo {author} {\bibfnamefont {H.-Y.}\ \bibnamefont {Lee}},
  \bibinfo {author} {\bibfnamefont {J.~H.}\ \bibnamefont {Han}},\ and\ \bibinfo
  {author} {\bibfnamefont {T.}~\bibnamefont {Senthil}},\ }\bibfield  {title}
  {\bibinfo {title} {Dipole condensates in tilted bose-hubbard chains},\ }\href
  {https://doi.org/10.1103/PhysRevB.107.195132} {\bibfield  {journal} {\bibinfo
   {journal} {Phys. Rev. B}\ }\textbf {\bibinfo {volume} {107}},\ \bibinfo
  {pages} {195132} (\bibinfo {year} {2023})}\BibitemShut {NoStop}%
\bibitem [{\citenamefont {Burchards}\ \emph {et~al.}(2022)\citenamefont
  {Burchards}, \citenamefont {Feldmeier}, \citenamefont {Schuckert},\ and\
  \citenamefont {Knap}}]{Burchards2022}%
  \BibitemOpen
  \bibfield  {author} {\bibinfo {author} {\bibfnamefont {A.~G.}\ \bibnamefont
  {Burchards}}, \bibinfo {author} {\bibfnamefont {J.}~\bibnamefont
  {Feldmeier}}, \bibinfo {author} {\bibfnamefont {A.}~\bibnamefont
  {Schuckert}},\ and\ \bibinfo {author} {\bibfnamefont {M.}~\bibnamefont
  {Knap}},\ }\bibfield  {title} {\bibinfo {title} {Coupled hydrodynamics in
  dipole-conserving quantum systems},\ }\href
  {https://doi.org/10.1103/PhysRevB.105.205127} {\bibfield  {journal} {\bibinfo
   {journal} {Phys. Rev. B}\ }\textbf {\bibinfo {volume} {105}},\ \bibinfo
  {pages} {205127} (\bibinfo {year} {2022})}\BibitemShut {NoStop}%
\bibitem [{\citenamefont {Morningstar}\ \emph {et~al.}(2023)\citenamefont
  {Morningstar}, \citenamefont {O'Dea},\ and\ \citenamefont
  {Richter}}]{Morningstar2023}%
  \BibitemOpen
  \bibfield  {author} {\bibinfo {author} {\bibfnamefont {A.}~\bibnamefont
  {Morningstar}}, \bibinfo {author} {\bibfnamefont {N.}~\bibnamefont {O'Dea}},\
  and\ \bibinfo {author} {\bibfnamefont {J.}~\bibnamefont {Richter}},\
  }\bibfield  {title} {\bibinfo {title} {Hydrodynamics in long-range
  interacting systems with center-of-mass conservation},\ }\href
  {https://doi.org/10.1103/PhysRevB.108.L020304} {\bibfield  {journal}
  {\bibinfo  {journal} {Phys. Rev. B}\ }\textbf {\bibinfo {volume} {108}},\
  \bibinfo {pages} {L020304} (\bibinfo {year} {2023})}\BibitemShut {NoStop}%
\bibitem [{\citenamefont {Gliozzi}\ \emph {et~al.}(2023)\citenamefont
  {Gliozzi}, \citenamefont {May-Mann}, \citenamefont {Hughes},\ and\
  \citenamefont {De~Tomasi}}]{Gliozzi2023}%
  \BibitemOpen
  \bibfield  {author} {\bibinfo {author} {\bibfnamefont {J.}~\bibnamefont
  {Gliozzi}}, \bibinfo {author} {\bibfnamefont {J.}~\bibnamefont {May-Mann}},
  \bibinfo {author} {\bibfnamefont {T.~L.}\ \bibnamefont {Hughes}},\ and\
  \bibinfo {author} {\bibfnamefont {G.}~\bibnamefont {De~Tomasi}},\ }\bibfield
  {title} {\bibinfo {title} {Hierarchical hydrodynamics in long-range
  multipole-conserving systems},\ }\href
  {https://doi.org/10.1103/PhysRevB.108.195106} {\bibfield  {journal} {\bibinfo
   {journal} {Phys. Rev. B}\ }\textbf {\bibinfo {volume} {108}},\ \bibinfo
  {pages} {195106} (\bibinfo {year} {2023})}\BibitemShut {NoStop}%
\bibitem [{\citenamefont {Ogunnaike}\ \emph {et~al.}(2023)\citenamefont
  {Ogunnaike}, \citenamefont {Feldmeier},\ and\ \citenamefont
  {Lee}}]{Ogunnaike_2023}%
  \BibitemOpen
  \bibfield  {author} {\bibinfo {author} {\bibfnamefont {O.}~\bibnamefont
  {Ogunnaike}}, \bibinfo {author} {\bibfnamefont {J.}~\bibnamefont
  {Feldmeier}},\ and\ \bibinfo {author} {\bibfnamefont {J.~Y.}\ \bibnamefont
  {Lee}},\ }\bibfield  {title} {\bibinfo {title} {Unifying emergent
  hydrodynamics and lindbladian low-energy spectra across symmetries,
  constraints, and long-range interactions},\ }\href
  {https://doi.org/10.1103/PhysRevLett.131.220403} {\bibfield  {journal}
  {\bibinfo  {journal} {Phys. Rev. Lett.}\ }\textbf {\bibinfo {volume} {131}},\
  \bibinfo {pages} {220403} (\bibinfo {year} {2023})}\BibitemShut {NoStop}%
\bibitem [{\citenamefont {Lux}\ \emph {et~al.}(2014)\citenamefont {Lux},
  \citenamefont {M\"uller}, \citenamefont {Mitra},\ and\ \citenamefont
  {Rosch}}]{Lux2014}%
  \BibitemOpen
  \bibfield  {author} {\bibinfo {author} {\bibfnamefont {J.}~\bibnamefont
  {Lux}}, \bibinfo {author} {\bibfnamefont {J.}~\bibnamefont {M\"uller}},
  \bibinfo {author} {\bibfnamefont {A.}~\bibnamefont {Mitra}},\ and\ \bibinfo
  {author} {\bibfnamefont {A.}~\bibnamefont {Rosch}},\ }\bibfield  {title}
  {\bibinfo {title} {Hydrodynamic long-time tails after a quantum quench},\
  }\href {https://doi.org/10.1103/PhysRevA.89.053608} {\bibfield  {journal}
  {\bibinfo  {journal} {Phys. Rev. A}\ }\textbf {\bibinfo {volume} {89}},\
  \bibinfo {pages} {053608} (\bibinfo {year} {2014})}\BibitemShut {NoStop}%
\bibitem [{\citenamefont {McCulloch}\ \emph {et~al.}(2023)\citenamefont
  {McCulloch}, \citenamefont {De~Nardis}, \citenamefont {Gopalakrishnan},\ and\
  \citenamefont {Vasseur}}]{McCullogh2023}%
  \BibitemOpen
  \bibfield  {author} {\bibinfo {author} {\bibfnamefont {E.}~\bibnamefont
  {McCulloch}}, \bibinfo {author} {\bibfnamefont {J.}~\bibnamefont
  {De~Nardis}}, \bibinfo {author} {\bibfnamefont {S.}~\bibnamefont
  {Gopalakrishnan}},\ and\ \bibinfo {author} {\bibfnamefont {R.}~\bibnamefont
  {Vasseur}},\ }\bibfield  {title} {\bibinfo {title} {Full counting statistics
  of charge in chaotic many-body quantum systems},\ }\href
  {https://doi.org/10.1103/PhysRevLett.131.210402} {\bibfield  {journal}
  {\bibinfo  {journal} {Phys. Rev. Lett.}\ }\textbf {\bibinfo {volume} {131}},\
  \bibinfo {pages} {210402} (\bibinfo {year} {2023})}\BibitemShut {NoStop}%
\bibitem [{\citenamefont {Wienand}\ \emph {et~al.}(2023)\citenamefont
  {Wienand}, \citenamefont {Karch}, \citenamefont {Impertro}, \citenamefont
  {Schweizer}, \citenamefont {McCulloch}, \citenamefont {Vasseur},
  \citenamefont {Gopalakrishnan}, \citenamefont {Aidelsburger},\ and\
  \citenamefont {Bloch}}]{wienand2023}%
  \BibitemOpen
  \bibfield  {author} {\bibinfo {author} {\bibfnamefont {J.~F.}\ \bibnamefont
  {Wienand}}, \bibinfo {author} {\bibfnamefont {S.}~\bibnamefont {Karch}},
  \bibinfo {author} {\bibfnamefont {A.}~\bibnamefont {Impertro}}, \bibinfo
  {author} {\bibfnamefont {C.}~\bibnamefont {Schweizer}}, \bibinfo {author}
  {\bibfnamefont {E.}~\bibnamefont {McCulloch}}, \bibinfo {author}
  {\bibfnamefont {R.}~\bibnamefont {Vasseur}}, \bibinfo {author} {\bibfnamefont
  {S.}~\bibnamefont {Gopalakrishnan}}, \bibinfo {author} {\bibfnamefont
  {M.}~\bibnamefont {Aidelsburger}},\ and\ \bibinfo {author} {\bibfnamefont
  {I.}~\bibnamefont {Bloch}},\ }\href@noop {} {\bibinfo {title} {Emergence of
  fluctuating hydrodynamics in chaotic quantum systems}} (\bibinfo {year}
  {2023}),\ \Eprint {https://arxiv.org/abs/2306.11457} {arXiv:2306.11457
  [cond-mat.quant-gas]} \BibitemShut {NoStop}%
\bibitem [{\citenamefont {Zechmann}\ \emph {et~al.}(2024)\citenamefont
  {Zechmann}, \citenamefont {Boesl}, \citenamefont {Feldmeier},\ and\
  \citenamefont {Knap}}]{zechmann2023dynamical}%
  \BibitemOpen
  \bibfield  {author} {\bibinfo {author} {\bibfnamefont {P.}~\bibnamefont
  {Zechmann}}, \bibinfo {author} {\bibfnamefont {J.}~\bibnamefont {Boesl}},
  \bibinfo {author} {\bibfnamefont {J.}~\bibnamefont {Feldmeier}},\ and\
  \bibinfo {author} {\bibfnamefont {M.}~\bibnamefont {Knap}},\ }\bibfield
  {title} {\bibinfo {title} {Dynamical spectral response of fractonic quantum
  matter},\ }\href {https://doi.org/10.1103/PhysRevB.109.125137} {\bibfield
  {journal} {\bibinfo  {journal} {Phys. Rev. B}\ }\textbf {\bibinfo {volume}
  {109}},\ \bibinfo {pages} {125137} (\bibinfo {year} {2024})}\BibitemShut
  {NoStop}%
\bibitem [{\citenamefont {Boesl}\ \emph {et~al.}(2024)\citenamefont {Boesl},
  \citenamefont {Zechmann}, \citenamefont {Feldmeier},\ and\ \citenamefont
  {Knap}}]{boesl2023}%
  \BibitemOpen
  \bibfield  {author} {\bibinfo {author} {\bibfnamefont {J.}~\bibnamefont
  {Boesl}}, \bibinfo {author} {\bibfnamefont {P.}~\bibnamefont {Zechmann}},
  \bibinfo {author} {\bibfnamefont {J.}~\bibnamefont {Feldmeier}},\ and\
  \bibinfo {author} {\bibfnamefont {M.}~\bibnamefont {Knap}},\ }\bibfield
  {title} {\bibinfo {title} {Deconfinement dynamics of fractons in tilted
  bose-hubbard chains},\ }\href
  {https://doi.org/10.1103/PhysRevLett.132.143401} {\bibfield  {journal}
  {\bibinfo  {journal} {Phys. Rev. Lett.}\ }\textbf {\bibinfo {volume} {132}},\
  \bibinfo {pages} {143401} (\bibinfo {year} {2024})}\BibitemShut {NoStop}%
\bibitem [{\citenamefont {Doshi}\ and\ \citenamefont
  {Gromov}(2021)}]{Doshi2021}%
  \BibitemOpen
  \bibfield  {author} {\bibinfo {author} {\bibfnamefont {D.}~\bibnamefont
  {Doshi}}\ and\ \bibinfo {author} {\bibfnamefont {A.}~\bibnamefont {Gromov}},\
  }\bibfield  {title} {\bibinfo {title} {{Vortices as fractons}},\ }\href
  {https://doi.org/10.1038/s42005-021-00540-4} {\bibfield  {journal} {\bibinfo
  {journal} {Commun. Phys.}\ }\textbf {\bibinfo {volume} {4}},\ \bibinfo
  {pages} {1} (\bibinfo {year} {2021})}\BibitemShut {NoStop}%
\bibitem [{\citenamefont {Zerba}\ \emph {et~al.}(2024)\citenamefont {Zerba},
  \citenamefont {Seidel}, \citenamefont {Pollmann},\ and\ \citenamefont
  {Knap}}]{zenodo}%
  \BibitemOpen
  \bibfield  {author} {\bibinfo {author} {\bibfnamefont {C.}~\bibnamefont
  {Zerba}}, \bibinfo {author} {\bibfnamefont {A.}~\bibnamefont {Seidel}},
  \bibinfo {author} {\bibfnamefont {F.}~\bibnamefont {Pollmann}},\ and\
  \bibinfo {author} {\bibfnamefont {M.}~\bibnamefont {Knap}},\ }\href
  {https://doi.org/10.5281/zenodo.13897005} {\bibinfo {title} {{Emergent
  Fracton Hydrodynamics in the Fractional Quantum Hall Regime of Ultracold
  Atoms}}} (\bibinfo {year} {2024})\BibitemShut {NoStop}%
\bibitem [{\citenamefont {Yao}\ and\ \citenamefont {et~al.}()}]{Yao2024}%
  \BibitemOpen
  \bibfield  {author} {\bibinfo {author} {\bibfnamefont {R.}~\bibnamefont
  {Yao}}\ and\ \bibinfo {author} {\bibnamefont {et~al.}},\ }\href@noop {}
  {\bibinfo {title} {in preparation and {B}ulletin of the {A}merican {P}hysical
  {S}ociety, {\bf{67}}, 7, {U}09.00008 (2022)}}\BibitemShut {NoStop}%
\bibitem [{\citenamefont {Girvin}\ \emph {et~al.}(1986)\citenamefont {Girvin},
  \citenamefont {MacDonald},\ and\ \citenamefont {Platzman}}]{Girvin1986}%
  \BibitemOpen
  \bibfield  {author} {\bibinfo {author} {\bibfnamefont {S.~M.}\ \bibnamefont
  {Girvin}}, \bibinfo {author} {\bibfnamefont {A.~H.}\ \bibnamefont
  {MacDonald}},\ and\ \bibinfo {author} {\bibfnamefont {P.~M.}\ \bibnamefont
  {Platzman}},\ }\bibfield  {title} {\bibinfo {title} {Magneto-roton theory of
  collective excitations in the fractional quantum hall effect},\ }\href
  {https://doi.org/10.1103/PhysRevB.33.2481} {\bibfield  {journal} {\bibinfo
  {journal} {Phys. Rev. B}\ }\textbf {\bibinfo {volume} {33}},\ \bibinfo
  {pages} {2481} (\bibinfo {year} {1986})}\BibitemShut {NoStop}%
\end{thebibliography}%
\end{document}